\documentclass{jfm}

\usepackage{graphicx}
\usepackage{amsmath,amssymb}
\usepackage{natbib}
\title[Energetics of collapsible channel flow with a nonlinear fluid-beam model]{Energetics of collapsible channel flow with a nonlinear fluid-beam model}
\author[D.~Y.~Wang, X.~Y.~Luo and P.~S.~Stewart]{D.~Y.~Wang, X.~Y.~Luo \and P.~S.~Stewart\footnote{peter.stewart@glasgow.ac.uk}}

\affiliation{School of Mathematics and Statistics, University of Glasgow, Glasgow, G12 8SQ UK }
\begin{document}
\maketitle

\begin{abstract}
We consider flow along a finite-length collapsible channel driven by a fixed upstream flux, where a section of one wall of a planar rigid channel is replaced by a plane-strain elastic beam subject to uniform external pressure. A modified constitutive law is used to ensure that the elastic beam is energetically conservative. We apply the finite element method to solve the fully nonlinear steady and unsteady systems. In line with previous studies, we show that the system always has at least one static solution and that there is a narrow region of the parameter space where the system simultaneously exhibits two stable static configurations: an (inflated) upper branch and a (collapsed) lower branch, connected by a pair of limit point bifurcations to an unstable intermediate branch. Both upper and lower static configurations can each become unstable to self-excited oscillations, initiating either side of the region with multiple static states. As the Reynolds number increases along the upper branch the oscillatory limit cycle persists into the region with multiple steady states, where interaction with the intermediate static branch suggests a nearby homoclinic orbit. These oscillations approach zero amplitude at the upper branch limit point, resulting in a stable tongue between the upper and lower branch oscillations. Furthermore, this new formulation allows us to calculate a detailed energy budget over a period of oscillation, where we show that both upper and lower branch instabilities require an increase in the work done by the upstream pressure to overcome the increased dissipation. 
\end{abstract}


\section{Introduction}

% physiological applications
There are many examples of collapsible tubes in the human body, including the blood vessels, airways and intestines. These vessels are typically sensitive to changes in internal pressure, so a variety of interesting physiological phenomena can arise when a flow is driven through them. For example, in veins above the heart the transmural (internal-external) pressure is often negative due to the hydrostatic pressure decrease with height, and so these vessels can spontaneously collapse \citep{moreno1970mechanics,wild1977viscous}. Furthermore, forced expiration of air from the lungs can induce airway collapse, significantly reducing the flow rate that can be expelled, in a phenomenon known as `flow limitation'  \citep{grotberg1989flutter}. This study is motivated by another physiological example, where the onset of self-excited oscillations in rapid flow along the brachial artery manifest as Korotkoff sounds during blood pressure measurement \citep{ur1970origin,bertram1989oscillations}. More expansive reviews of the interesting phenomena that can arise from flow through collapsible vessels are provided elsewhere \citep{Heil&Hazel2011,Grotberg&Jensen2004}.

% experiments
Flow in collapsible tubes can be investigated experimentally using a Starling Resistor, where liquid is driven through a segment of externally pressurised flexible tubing mounted between two rigid tubes \cite[e.g.][]{bertram1982mathematical,bertram1986unstable,bertram1990mapping,bertram1991application,bertram2006onset}. A typical experiment drives flow along the tube with either a fixed upstream flow rate or a fixed upstream pressure. The experiments demonstrate that the system can adopt a steady configuration with a non-uniform wall profile, while for some operating conditions the system can even exhibit multiple (stable) steady states in a so-called `open-to-closed' transition  \citep{bertram1991application}. On top of this static behaviour, a wide variety of different classes of self-excited oscillation have been observed. These oscillations fall into distinct frequency bands \citep{bertram1990mapping} and exhibit complex nonlinear limit cycles, characterised by phase portraits of quantities such as pressure or flow rate at different points along the tube \citep{bertram1990mapping,bertram1991application}. However, the mechanisms which generate these oscillations are still not well understood. 

% ODE structure
Theoretical study of the Starling Resistor experiment began with lumped parameter models \citep[eg][]{katz1969flow,bertram1982mathematical}, formed from a small number of ordinary differential equations. Such models qualitatively capture the static deformation of the tube observed in experiments, predicting that over a narrow window of the parameter space the system can exhibit three co-existing steady states: an upper branch (where the wall is inflated), a lower branch (where the wall is collapsed) connected by an (unstable) intermediate branch by a pair of limit point bifurcations \citep{armitstead1996study}. These lumped models further predict the onset of self-excited oscillations from the collapsed (lower) static branch \citep{bertram1982mathematical} and elucidate possible global interactions between the oscillatory limit cycles and the additional static solutions \citep{armitstead1996study}.

% 3d collapsible tubes
In more recent years these theoretical studies have gradually increased in complexity, beginning with cross-sectionally averaged one-dimensional models \citep[eg][]{cancelli1985separated,jensen1990instabilities} all the way to full three-dimensional models of the static \citep{hazel2003steady,marzo2005three,zhang2018three} and oscillatory behaviour \citep[][]{heil1997stokes,heil2010self,whittaker2010predicting} of the tube. However, numerical solutions of the latter requires immense computational resources and so it has not yet been possible to fully map out the different classes of oscillatory behaviour across the parameter space.

% channel analogue
For simplicity in probing the underlying mechanisms of instability, theoretical attention has also focused on a planar analog of the Starling Resistor setup, formed by removing a segment of one wall of a rigid channel and replacing by an externally pressurised flexible membrane \citep{pedley1992longitudinal}. This channel system has subsequently been studied using fully nonlinear simulations \citep{luo1995numerical,luo1996numerical,jensen2003high,xu2014resonance}, where many of the predictions are analogous to those from the lumped parameter models. For example, for the flexible wall modelled as either a thin membrane or a nonlinearly elastic shell, the system can exhibit multiple co-existing steady states provided the Reynolds number is large enough to collapse the channel through the Bernoulli effect \citep{luo2000multiple,heil2004efficient}. Furthermore, the lower branch of static solutions is unstable to oscillations when the channel becomes sufficiently collapsed \citep{heil2004efficient}. In addition, recent work by \cite{herrada2021global} has shown that the upper branch of static solutions can also become unstable to oscillations provided the external pressure is sufficiently low (so the flexible wall bulges outwards). 

% low order models
Further insight into the mechanisms of instability has been provided by spatially one-dimensional models of the collapsible channel system, derived using a flow profile assumption \citep{stewart2009local,xu2013divergence}. Using the approach, \citet{stewart2017instabilities} described two families of self-excited oscillations arising from the lower branch of static solutions in the limit of large external pressure, where the primary global instability is to a low-frequency mode from a static state which is inflated along most of the compliant segment, but collapsed along a narrow layer adjacent to the downstream rigid segment. Similarly, \citet{xu2013divergence} considered a similar system with an imposed gradient in external pressure on the flexible segment (which ensures that the flat wall is always a steady state of the system); this results in a rather different static structure to those reported with constant external pressure, where non-trivial static modes emerge via a transcritical bifurcation in the inviscid limit at particular values of the pre-stress parameter. Self-excited oscillations then arise as a resonance between two modes, which can be explored in the limit of large downstream channel length \citep{xu2014resonance}, where the oscillations eventually grow to exhibit vigorous `slamming' oscillations \citep{xu2015low}.

% collapsible channel with beam
This study focuses on the planar channel analog of the Starling Resistor, but instead models the elastic wall as a plane-strained nonlinear beam that is materially linear but geometrically nonlinear \citep{gere2003mechanics}; this beam has resistance to both bending and stretching. This model was first introduced by \citet{cai2003fluid} and has subsequently been explored in depth by Luo and coworkers, examining flow driven by either prescribed upstream flux \citep{luo2008cascade,hao2016arnoldi} or prescribed upstream pressure \citep{liu2012stability,hao2016arnoldi}. In this study we focus on the flow driven system, where it has been shown that the system admits multiple families of self-excited oscillations in a novel cascade structure: the system has isolated regions of stability between unstable regions which each correspond to a different number of extrema in the wall profile \citep{luo2008cascade}. Their neutral stability curve, plotted in the parameter space spanned by the Reynolds number and the dimensionless resistance to beam stretching ($c_\lambda$), is plotted in figure \ref{Fig:st-saddle-neutral} below. In this paper we revisit the model of \citet{luo2008cascade}, making a small correction to the constitutive law for the beam to ensure that it is energetically conservative. We explore the structure of the underling static solutions and show that this is analogous to those described above, with regions of parameter space which exhibit three co-existing steady states. We examine the stability of these static states using fully nonlinear simulations, predicting an additional mode of oscillation not noted by \citet{luo2008cascade}, arising as an instability of the upper branch of static solutions \citep[analogous to upper branch instability recently reported by][]{herrada2021global}. We investigate the saturated limit cycles of this oscillatory mode and its subsequent interaction with the other static solutions. 

% energy budget including forced oscillations
Analysing the energy budget of the flow has proved to be a useful tool for probing the mechanism of self-excited oscillations, particularly for oscillations driven by a prescribed upstream pressure \citep{jensen2003high,stewart2009local,stewart2010sloshing}. The approach is similar to the classical derivation of the Reynolds--Orr equation \citep{schmid2001stability}, where we take the scalar product between the fluid momentum equations and the corresponding fluid velocity and integrate over the channel area to construct a rate of energy balance; for incompressible flow the terms appearing in this energy rate equation can be written in the form
\begin{equation}
{\cal K} + {\cal E} = {\cal F} + {\cal P} - {\cal D},
\end{equation}
where ${\cal K}$ is the rate of change of kinetic energy, ${\cal E}$ is the rate at which fluid does work on the flexible wall, ${\cal F}$ is the net rate of energy extraction from the mean flow, ${\cal P}$ is the rate of working of the upstream driving pressure and ${\cal D}$ is the rate of working of viscous dissipation. The overall energy change arising from each term can be obtained by taking the time average over a period of fully developed oscillation; throughout this paper we denote this integral with the superscript $^{(avg)}$. For a conservative wall model (where no work is done on the wall, so ${\cal E}^{(avg)}=0$), the time-averaged energy budget can be expressed as the sum of two sources, the time-averaged work done by the upstream driving pressure (${\cal P}^{(avg)}$) and the time-averaged energy flux extracted from the mean flow (${\cal F}^{(avg)}$), balanced by the energy consumed by the time-averaged working of viscous dissipation (${\cal D}^{(avg)}$), so that
\begin{equation}
{\cal F}^{(avg)}+{\cal P}^{(avg)} = {\cal D}^{(avg)}.
\end{equation}
The net energy flux ${\cal F}^{(avg)}$ is extremely important for some flows driven by upstream pressure. For example, for `sloshing' oscillations, found in both flexible-walled channels \citep{jensen2003high,stewart2009local,stewart2010sloshing} and collapsible tubes \citep{whittaker2010predicting}, it has been shown that in in the limit of large wall pre-stress the time-averaged dissipation can be decomposed into the sum of two components, the energy lost due to dissipation in the mean flow (denoted ${\cal D}^{(avg)}_P$) and the energy lost due to dissipation in the oscillation (denoted ${\cal D}^{(avg)}_S$). It emerges that exactly two-thirds of the energy flux is consumed by dissipative effects in the oscillation (${\cal D}_S^{(avg)}= \tfrac{2}{3}{\cal F}^{(avg)}$), while the remaining one-third increases the mean flow. However, there is evidence to suggest that ${\cal F}^{(avg)}$ may become negative for lower pre-stress \citep{stewart2010sloshing}, indicating a different mechanism entirely. In this study we explore the mechanism of energy transfer in flux-driven oscillations and show that ${\cal F}^{(avg)}$ plays no role in the mechanism of instability.

% outline of the paper
In this paper we revisit the nonlinear fluid-beam model presented by \citet{luo2008cascade}, reformulating the nonlinear governing equations to derive the full nonlinear energy budget for oscillatory limit cycles (\S\ref{sec: method}), adjusting the Kirchhoff constitutive law for the beam to ensure the elastic wall is energetically conservative. In \S \ref{sec: B_NumResults} we examine the behaviour of our new fluid-beam model along two slices of the parameter space (shown on figure \ref{Fig:st-saddle-neutral}), characterising both the steady (\S \ref{sec:st}) and unsteady (\S\ref{sec:unst}) behaviour of the system and provide an overview of the parameter space for fixed external pressure (\S \ref{sec:overview}). In particular, we elucidate an instability of the upper branch of static solutions (\S \ref{sec:upper}) which is similar to that found by \citet{herrada2021global}, explore the nonlinear bifurcation structure of the system (\S \ref{sec:bifurcation}), the resulting interaction between the oscillatory limit cycles and the folding point associated with the upper branch of static solutions (\S \ref{sec:homoclinic}) and the energy budget of self-excited oscillations (\S \ref{sec:energy}). Finally, in \S\ref{sec:compare} we compare the predictions of our new Kirchhoff law to the previous law used by \citet{luo2008cascade}, showing that the change makes very little difference to the predictions. 

\section{The model}
\label{sec: method}

\begin{figure}
\centering
\includegraphics[width=0.95\textwidth]{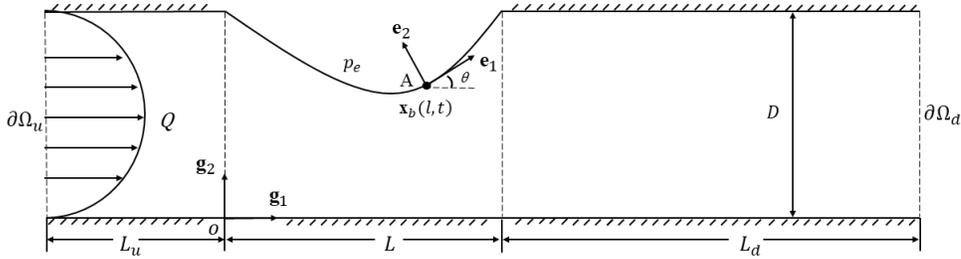}
\caption{Schematic of the fluid-beam model.}
\label{Fig:BemMod}
\end{figure}
	
We consider a planar rigid channel of finite length $L_0$ and uniform width $D$ containing a viscous fluid. An internal segment of length $L$ of one wall is replaced by a planar elastic beam in a state of plane strain. The corresponding lengths of the upstream and downstream rigid segments are denoted $L_u$ and $L_d$, respectively, so $L_0=L_u+L+L_d$. This elastic beam is initially of uniform thickness $h$, which we assume to be much less than the channel width ($h \ll D)$. We denote the two-dimensional fluid domain by $\Omega$ and use $\partial\Omega_u$, $\partial\Omega_d$, $\partial\Omega_b$ to denote the upstream fluid inlet, the downstream fluid outlet, and the elastic beam, respectively. This setup is similar to the system introduced by \citet{pedley1992longitudinal} and has been analysed extensively by \citet{cai2003fluid} and \citet{luo2008cascade}.

% fluid parameters
We consider a parabolic inlet flow to the channel with flux $Q$ (per unit width in the out-of-plane direction) with average velocity $Q/D$. We choose the outlet pressure along $\partial\Omega_d$ to be zero, our pressure to which all other pressures and stresses are compared. The fluid is assumed to be Newtonian and incompressible with constant density $\rho$ and viscosity $\mu$. The elastic beam is subject to a uniform external pressure, denoted $p_e$. 

We establish two coordinate systems to describe the motion, as shown in figure \ref{Fig:BemMod}. Firstly, ${\mathbf g}_1,~{\mathbf g}_2,~{\mathbf g}_3$ are the unit vectors of the Cartesian coordinate system in the (undeformed) reference configuration, such that ${\mathbf g}_1$ is oriented along the entirely rigid wall, ${\mathbf g}_2$ is normal to $\mathbf g_1$ in the plane of the channel (pointing into the channel) and ${\mathbf g}_3$ is in the out-of-plane direction. Conversely, $\mathbf e_1,~\mathbf e_2,~\mathbf e_3$ are unit vectors of the material coordinates in the (deformed) current configuration of the beam, where $\mathbf e_1$ is the local tangent to the beam, $\mathbf e_2$ is the local normal to the beam (pointing out of the channel),  and $\mathbf e_3$ is normal to the plane of the channel ($\mathbf g_3=\mathbf e_3$). In what follows, we ignore deflections in the out-of-plane direction and so consider all vectors as two-dimensional.

In the absence of fluid loading we assume the elastic beam is flat and parallel to the entirely rigid wall. Following \citet{liu2012stability}, we consider a massless elastic beam and denote the axial pre-stress along the beam as $T$. Further, we denote $EA$ and $EJ$ as the extensional stiffness and bending stiffness of the beam, respectively, where $E$ is the Young's modulus of the material while $A$ and $J$ are the cross-sectional area and the second moment of inertia of the cross-sectional area of the beam with respect to the ${\mathbf g}_1$ axis, respectively.
 
We denote an arbitrary point on the (flat) beam in the reference configuration as
\begin{equation}
{\mathbf X}_b(l)=l\mathbf g_1+D\mathbf g_2,~~~~(0 \le l \le L). 
\end{equation}
After deformation this point moves to,  
\begin{equation}
\mathbf x_b(l,t)=x_b(l,t)\mathbf g_1+y_b(l,t)\mathbf g_2,~~~~(0\le l \le L), 
\end{equation}
where we use the subscript $b$ to denote points on the beam. The principal stretch of the beam is then
\begin{equation}
\lambda=\left[{\left(\frac{\partial x_b}{\partial l}\right)^2+\left(\frac{\partial y_b}{\partial l}\right)^2}\right]^{\frac{1}{2}},~~~~(0\le l\le L).
\end{equation}
We denote $\theta$ as the angle between the tangent to the deformed beam $\mathbf e_1$ and the unit vector ${\mathbf g_1}$ (see figure \ref{Fig:BemMod}), hence we have
\begin{equation}
\frac{\partial x_b}{\partial l}=\lambda\cos\theta,\quad\frac{\partial y_b}{\partial l}=\lambda\sin\theta.\label{eq:theta}
\end{equation}
The arc-length coordinate, denoted $s$, is measured along the beam from the upstream intersection with the rigid wall, computed as 
\begin{equation}
s(l,t)=\int_0^l\lambda(l',t)\,{\rm d}l',~~~~(0\le l\le L).
\end{equation}
The total length of the deformed beam is therefore $S=s(L,t)$.

\subsection{Governing equations}
\label{sec: goveq}

We introduce dimensionless variables by scaling all lengths on the channel width $D$, velocities on the mean inlet flow $Q/D$, time on $D^2/Q$ and pressures on the inertial pressure scale $\rho Q^2/D^2$ (where the fluid outlet pressure is set to zero without loss of generality). Under the scaling we obtain four dimensionless parameters associated with the geometry of the channel in the form
\begin{equation}
\tilde L_u=\frac{L_u}{D}, \quad \tilde L_d=\frac{L_d}{D}, \quad \tilde L=\frac{L}{D}, \quad \tilde h=\frac{h}{D},
\end{equation}
the dimensionless lengths of the upstream, downstream and collapsible segments of the channel and the dimensionless beam thickness, respectively. We also obtain three dimensionless parameters associated with the elasticity of the beam, in the form
\begin{equation}
\tilde c_\lambda=\frac{(EA)D}{\rho Q^2 }, \quad \tilde c_\kappa=\frac{EJ}{\rho Q^2 D}, \quad \tilde T=\frac{TD}{\rho Q^2}, 
\end{equation}
the dimensionless extensional and bending stiffnesses and the dimensionless beam pre-tension, respectively. Finally, the flow is also governed by the Reynolds number and the dimensionless external pressure, which take the form
\begin{equation}
\tilde Re=\frac{Q\rho}{\mu}, \quad \tilde{p}_e=\frac{p_e D^2}{\rho Q^2}.
\end{equation}
We henceforth focus on the dimensionless quantities and drop the tildes for simplicity. 

The governing equations for the (two-dimensional) fluid velocity ${\mathbf u}({\mathbf x},t)$ and pressure $p({\mathbf x},t)$ follow from the incompressible Navier-Stokes equations in the form,
\begin{equation}
\nabla\cdot{\mathbf u}=0,\quad\frac{\partial{\mathbf u}}{\partial t}+\left({\mathbf u}\cdot\nabla\right){\mathbf u}=\nabla\cdot{\boldsymbol{\sigma}}, ~~~~({\mathbf x}\in\Omega).\label{FGvEq}
\end{equation}
For the Newtonian fluid we have 
\begin{equation}
{\boldsymbol{\sigma}}=-p{\mathbf I}+Re^{-1}\left(\nabla{\mathbf u}+{\nabla{\mathbf u}}^{\rm{T}}\right),~~~~({\mathbf x}\in\Omega),\label{eq:sigma}
\end{equation}
where  $\mathbf I$ is the identity matrix and the superscript $^{\rm{T}}$ represents the matrix transpose. 

To establish governing equations for the massless beam we consider a virtual displacement of a differential element and impose conservation of linear and angular momentum, following the derivation of \citet{cai2003fluid}. We denote the internal force acting on a cross-section of the beam as $\mathbf F=F_1{\mathbf e_1}+F_2{\mathbf e_2}$ and the external force acting on the beam as $\mathbf q ={\sigma}_1{\mathbf e}_1+({\sigma}_2-p_e){\mathbf e}_2$, where ${\sigma_1}$ and  ${\sigma_2}$ are the tangent and normal components of the fluid traction on the beam,
\begin{equation}
\sigma_1=\left(-{\boldsymbol\sigma}{\mathbf e_2}\right)\cdot{\mathbf e_1},\quad\sigma_2=\left(-{\boldsymbol\sigma}{\mathbf e_2}\right)\cdot{\mathbf e_2}. \end{equation}
Denoting $\partial f$ as a virtual displacement of the variable $f$, conservation of linear momentum takes the form 
\begin{equation}
{\partial }\mathbf F+\mathbf q{\partial }s={\mathbf 0},~~~~({\mathbf x}\in\partial\Omega_b).\label{BeMomEq}
\end{equation}
Similarly, denoting $\mathbf M=M{\mathbf e_3}$ as the moment acting on the beam, conservation of angular momentum for the beam element takes the form
\begin{equation}
{\partial }\mathbf M+\mathbf x_b\times{\mathbf q} \partial s+(\mathbf x_b+\partial \mathbf x_b) \times(\mathbf F+\partial \mathbf F)-\mathbf x_b\times\mathbf F={\mathbf 0},~~~~({\mathbf x}\in\partial\Omega_b).\label{BeAngMomEq}
\end{equation}
The commonly used linear constitutive laws for elastic beams \citep{gere2003mechanics}, which were adopted in the previous fluid-beam model \citep{cai2003fluid}, take the form
\begin{equation}
F_1=T+c_{\lambda}(\lambda-1),\quad M=c_{\kappa}\kappa,\label{KirLaw-old}
\end{equation}
where $\kappa$ is the dimensionless curvature of the beam defined as 
\begin{equation}
\kappa=\frac{\partial\theta}{\partial s}. 
%\frac{1}{\lambda^3}\left(\frac{\partial x_b}{\partial l}\frac{\partial ^2y_b}{\partial l^2}-\frac{\partial y_b}{\partial l}\frac{\partial^2x_b}{\partial l^2}\right).
\label{Curvature: Curr}
\end{equation}
However, to ensure that our approach is suitable to describe large deformations we instead formulate these constitutive laws with respect to the reference configuration of the beam (parameterised by $l$) and we replace $(\ref{KirLaw-old})$ with a modified form
\begin{equation}
F_1=T+c_{\lambda}(\lambda-1),\quad M=c_{\kappa} \frac{\partial\theta}{\partial l},
\label{KirLaw}
\end{equation}
where we can also prove the identity \citep{wang2019energetics} 
\begin{equation}
\frac{\partial\theta}{\partial l}=\lambda\kappa.
\label{BGvEq:thetalamkapa}
\end{equation}
These new expressions (\ref{KirLaw}) reduce to (\ref{KirLaw-old}) in the limit of small displacements. The advantages of introducing these constitutive laws will become clearer below. Using either of these two constitutive laws, we can eliminate the unknown $F_2$ between the equations (\ref{BeMomEq}) and (\ref{BeAngMomEq}) and end up with a closed system.

The dimensionless governing equations for the beam can be written as
\begin{eqnarray}
c_{\kappa}\kappa\frac{\partial \left(\lambda\kappa\right)}{\partial l}+c_{\lambda}\frac{\partial \lambda}{\partial l}+\lambda{\sigma_1}&=0,\label{lBGvEq:1}\\
-c_{\kappa}\frac{\partial }{\partial l}\left(\frac{1}{\lambda}\frac{\partial(\lambda\kappa)}{\partial l}\right)+c_{\lambda}\lambda\kappa(\lambda-1)+\lambda\kappa T+\lambda{\sigma_2}-\lambda p_e&=0.\label{lBGvEq:2}
\end{eqnarray}

For flow boundary conditions along the channel inlet ($\partial \Omega_u$) we prescribe a parabolic inlet flow with unit flux in the form ${\mathbf u}=6y(1-y){\mathbf g}_1$. Along the channel outlet we impose a stress free condition in the form  ${\boldsymbol{\sigma}}{\mathbf g}_2={\mathbf 0}$. Note that this is not formally identical to imposing zero pressure along the outlet, but we assume that the downstream rigid segment is sufficiently long so that the outflow is approximately parallel and the normal viscous stress terms are negligible \cite[a similar approach was used by][]{jensen2003high}.  We assume the no-slip condition along the rigid walls as well as continuity of velocity between the elastic beam and the fluid in the form
\begin{eqnarray}
{\mathbf u}&={\mathbf 0},~~\quad&(y=0;~~y=1, -L_u\leq x\leq 0,~ L\leq x\leq L+L_d),\label{FluBoCon:1}\\
{\mathbf u}&={\mathbf u_{b}},\quad&({\mathbf x}\in\partial\Omega_b).\label{FluBoCon:2}
\end{eqnarray}
The two ends of the elastic beam are attached to the rigid wall at a fixed angle, in the form
\begin{align}
x_b(0,t)=0,\quad y_b(0,t)=1,
\quad x_b(L,t)=L,\quad y_b(L,t)=1,\quad \theta(0,t)=\theta(L,t)=0.\label{BBodCon:1}
\end{align}

\subsection{Fully nonlinear energy budget}
\label{sec: full ener}

A useful tool for analysing the mechanism of instability in collapsible channel flows is to formulate the energy budget of the system \citep[e.g.][]{jensen2003high, stewart2009local, stewart2010sloshing}.  Here we perform the energy budget analysis for our updated fluid-beam model. To formulate the energy equation we begin with the fluid and consider the dot product of the fluid velocity with the fluid momentum equations (\ref{FGvEq}). As the fluid is incompressible, we manipulate and integrate this energy equation over the fluid domain $\Omega$ to obtain the total energy budget of the system in the form
\begin{align}
&\int_{\Omega}\tfrac{1}{2}\frac{\partial({\mathbf u}\cdot{\mathbf u})}{\partial t}\,{ \rm d}A+\int_{\Omega}\tfrac{1}{2}\nabla\cdot\left(({\mathbf u}\cdot{\mathbf u}){\mathbf u}\right)\,{\rm d}A\notag\\
&\indent=\int_{\Omega}\nabla\cdot(-p{\mathbf u})\,{\rm d}A+\int_{\Omega}\frac{1}{Re}\left[\nabla\cdot\left((\nabla{\mathbf u}+\nabla{\mathbf u}^{\text{T}}){\mathbf u}\right)-\text{Tr}\left((\nabla{\mathbf u}+\nabla{\mathbf u}^{\text{T}})\nabla{\mathbf u}\right)\right]\,{\rm d}A.\label{FEnerEq}
\end{align}
By the Reynolds transport theorem and the divergence theorem, this energy budget can be rearranged as (see \citet{wang2019energetics} for details)
\begin{align}
&\frac{\partial}{\partial t}\int_{\Omega}\tfrac{1}{2}{\mathbf u}\cdot{\mathbf u}\,{\rm d}A+\left[\int_0^1\tfrac{1}{2}({\mathbf u}\cdot{\mathbf u})\left({\mathbf u}\cdot{\mathbf g_1}\right)\,{\rm d}y\right]_{x=-L_u}^{x=L+L_d}\notag\\
&\indent=\left[\int_0^1-p{\mathbf u}\cdot{\mathbf g_1}\,{\rm d}y\right]^{x=L+L_d}_{x=-L_u}
-\frac{1}{Re}\int_{\Omega}\text{Tr}\Big(\left(\nabla{\mathbf u}+\nabla{\mathbf u}^{\text{T}}\right)\nabla{\mathbf u}\Big)\,{\rm d}A\notag\\
&\quad \quad +\int_{\partial\Omega_b} \left\{ -p+\frac{1}{Re}\left(\nabla{\mathbf u}+\nabla{\mathbf u}^\text{T}\right)\right\}{\mathbf u}\cdot{\mathbf e_2}\,{\rm d}s,
\end{align}
where the final term involves the surface integral of the fluid stress along the deformed beam, parametrised by $s$. Applying the definition of the fluid stress tensor (\ref{eq:sigma}) to the integral evaluated on the elastic beam, we write the total energy budget of the system as
\begin{equation}
{\cal P}+{\cal F}-{\cal D}={\cal K}+{\cal E},\label{FlEnerBug}
\end{equation}
where
\begin{align}
{\cal P}&=-\left[\int_0^1p{\mathbf u}\cdot{\mathbf g_1}\,{\rm d}y\right]^{x=L+L_d}_{x=-L_u},\\
{\cal F}&=-\left[\int_0^1\tfrac{1}{2}({\mathbf u}\cdot{\mathbf u})\left({\mathbf u}\cdot{\mathbf g_1}\right)\,{\rm d}y\right]_{x=-L_u}^{x=L+L_d},\\
{\cal K}&=\frac{\partial}{\partial t}\int_{\Omega}\tfrac{1}{2}{\mathbf u}\cdot{\mathbf u}\,{\rm d}A,\label{FulFluEnerBud:1}\\
{\cal E}&=
-\int_{\partial\Omega_b}({\boldsymbol\sigma}{\mathbf e_2})\cdot{\mathbf u}\,{\rm d}s,\label{FluEnerBug:Ef}\\
{\cal D}&=\frac{1}{Re}\int_{\Omega}\text{Tr}\Big(\left(\nabla{\mathbf u}+\nabla{\mathbf u}^{\text{T}}\right)\nabla{\mathbf u}\Big)\,{\rm d}A.
\end{align}
Here ${\cal P}$ is the rate of working of pressure  at the channel inlet (since $p$ is set to zero along the channel outlet), ${\cal F}$ is the net kinetic energy flux extracted from the mean flow between the channel ends, ${\cal K}$ is the rate of working of kinetic energy, ${\cal E}$ is the rate of working of fluid stress on the beam and ${\cal D}$ is the rate of dissipative energy loss due to fluid viscosity. This dissipation of energy is non-negative, since ${\cal D}$ can be alternatively expressed in terms of the velocity components ${\mathbf u}=u_1{\mathbf g_1}+u_2{\mathbf g_2}$ as
\begin{align}
{\cal D}&=\frac{1}{Re}\int_{\Omega}\left[2\left(\frac{\partial u_1}{\partial x}\right)^2+2\left(\frac{\partial u_2}{\partial y}\right)^2+\left(\frac{\partial u_1}{\partial y}+\frac{\partial u_2}{\partial x}\right)^2\right]\,{\rm d}A.\label{FulFluEnerBud:5}
\end{align}
Note that this formulation represents an improvement on the energy budget presented by \citet{stewart2010sloshing}, since the dissipation term they derived included part of the work done by viscous stresses on the wall and so could take either sign depending on the parameters.

To fully evaluate the work done by the fluid on the wall (${\cal E}$) we substitute the beam equations (\ref{lBGvEq:1}, \ref{lBGvEq:2}), the identity (\ref{BGvEq:thetalamkapa}) and boundary conditions (\ref{FluBoCon:1}-\ref{BBodCon:1}) into (\ref{FluEnerBug:Ef}) to obtain
\begin{align}
{\cal E}={\cal U}_{\kappa}+{\cal U}_{\lambda}-{\cal P}_e,\label{FEnerEx}
\end{align}
where,
\begin{align}
&{\cal U}_\kappa=\frac{\partial}{\partial t}\int_0^{L}\tfrac{1}{2}c_\kappa(\lambda\kappa)^2\,{\rm d}l, \label{kappa}\\
&{\cal U}_\lambda=\frac{\partial}{\partial t}\int_0^{L}\left[T(\lambda-1)+\tfrac{1}{2}c_\lambda\left(\lambda-1\right)^2\right]\,{\rm d}l, \\
&{\cal P}_e=\int_0^{L}-p_e\Big({\mathbf u_b}\cdot{\mathbf e_2}\Big)\lambda\,{\rm d}l=\frac{\partial}{\partial t}\int_0^{L}-p_e\frac{\partial x_b}{\partial l}y_b\,{\rm d}l. 
\label{eq:extpress}
\end{align}
Here ${\cal U}_{\kappa}$ is the rate of working of bending stresses, ${\cal U}_{\lambda}$ is the rate of working of extensional stresses and ${\cal P}_e$ is the rate of working of external pressure. Note that the rate of working of external pressure ${\cal P}_e$ has been manipulated using the boundary condition (\ref{BBodCon:1}); for details see Appendix \ref{appA}.

Using the fluid-beam model proposed by \citet{cai2003fluid} with constitutive law (\ref{KirLaw-old}), \citet{liu2012stability} established a similar energy budget, but in this case the rate of working of bending stresses (${\cal U}_{\kappa}$) could not be written as a complete time derivative and so their system is not energetically conservative when averaged over a period of self-excited oscillation. On the other hand,  applying the nonlinear constitutive law (\ref{KirLaw}) for the beam makes the system energetically conservative. This difference will also result in minor changes to the other energy terms in (\ref{FEnerEx}). 

We now summarise the fully non-linear energy budget for the coupled fluid-beam system in the form 
\begin{align}
{\cal P}+{\cal F}+{\cal P}_e={\cal K}+{\cal D}+{\cal U}_{\kappa}+{\cal U}_{\lambda},
\label{FulSysEnerBug:0}
\end{align}
where the left-hand side represents the sources of energy into the system and the right-hand side represents the losses of energy.

\subsection{Energetics of steady flow}
\label{sec:st_energy}

In the steady state, in which all variables are independent of time, we denote the flow velocity and pressure as ${\mathbf u}^{(0)}$ and $p^{(0)}$, respectively. In this case the terms ${\cal P}_e$, ${\cal K}$, ${\cal U}_{\kappa}$ and ${\cal U}_{\lambda}$ all vanish. Therefore, the energy budget for the steady system can be expressed simply as,
\begin{align}
{\cal P}^{(0)}+{\cal F}^{(0)}={\cal D}^{(0)},
\label{B_FullEner:st}
\end{align}
with
\begin{align}
{\cal F}^{(0)}&=-\left[\int_0^1\left(\tfrac{1}{2}\left({\mathbf u}^{(0)}\cdot{\mathbf u}^{(0)}\right)\left({\mathbf u}^{(0)}\cdot{\mathbf g}_1\right)\right)\,{\rm d}y\right]_{x=-L_u}^{x=L+L_d},\\
{\cal P}^{(0)}&=-\left[\int_0^1\left(p^{(0)}{\mathbf u}^{(0)}\cdot{\mathbf g}_1\right)\,{\rm d}y\right]^{x=L+L_d}_{x=-L_u},\\
{\cal D}^{(0)}&=\frac{1}{Re}\int_{\Omega^{(0)}}\text{Tr}\left(\left(\nabla{\mathbf u}^{(0)}+\left(\nabla{\mathbf u}^{(0)}\right)^{\text{T}}\right)\nabla{\mathbf u}^{(0)}\right)\,{\rm d}A,\label{B_FullEner:st_3}
\end{align}
where $\Omega^{(0)}$ denotes the fluid domain in the steady state. 

\subsection{Energetics of fully developed oscillations}
\label{sec:avg_energy}

For an unsteady oscillation which has saturated into a nonlinear (finite-amplitude) limit cycle (see \S\ref{sec:unst})  we average over a period of oscillation. For example for quantity $f(t)$, we compute the time average as
\begin{align}
\label{eq:timeaverage}
f^{(avg)}=\frac{1}{\tau}\int_t^{t+\tau}f(t')\,{\rm}dt',
\end{align}
where $\tau$ is the period of oscillation. In this case, the rate of working of external pressure, ${\cal P}_e^{(avg)}$, the average rate of working of fluid kinetic energy, ${\cal K}^{(avg)}$, and the average of the rate of working of bending and extensional stiffness, ${\cal U}_{\kappa}^{(avg)}$ and  ${\cal U}_{\lambda}^{(avg)}$, all vanish. Therefore, the energy budget of the unsteady system (\ref{FulSysEnerBug:0}) averaged over one period of oscillation becomes simply
\begin{align}
{\cal P}^{(avg)}+{\cal F}^{(avg)}={\cal D}^{(avg)},
\end{align}
where ${\cal P}^{(avg)}$ denotes the average rate of working of the upstream pressure over a period, ${\cal F}^{(avg)}$ is the average of the kinetic energy flux extracted from the mean flow over a period and ${\cal D}^{(avg)}$ denotes the average rate of energy loss due to viscosity over a period.

In analysing self-excited oscillations, it is useful to consider the excess energy due to oscillation by subtracting the corresponding steady components in the form  
\begin{align}
{\cal P}^{(e)}={\cal P}^{(avg)}-{\cal P}^{(0)},\quad {\cal F}^{(e)}={\cal F}^{(avg)}-{\cal F}^{(0)},\quad {\cal D}^{(e)}={\cal D}^{(avg)}-{\cal D}^{(0)}.
\label{B_FullEner:Osci}
\end{align}
These energy terms are all computed in \S\ref{sec:energy} below.

\subsection{The finite element method}
\label{sec:B_NumMethod}

A finite-element method is used to solve the coupled beam-fluid system and construct the corresponding energy budget.
We divide the fluid domain into three sections, %shown in \textcolor{red}{figure \ref{Fig:BemMod}}
denoted as A, B and C for the upstream, compliant and downstream compartments, respectively, as described in \citet{luo1996numerical} \citep[see also][]{luo2008cascade}.
We use an adaptive mesh in section B (where the beam is deformable), while we use a fixed mesh for sections A and C. The flow is described using an Eulerian description for sections A and C, whereas the flow is described using an Arbitrary Lagrangian-Eulerian (ALE) method for section B \citep{donea1982arbitrary}. 

Across the elastic segment (section B) we employ rotating spines following  \citet{cai2003fluid}, where we seed nodes along the rigid wall and connect these to nodes on the beam by spines; nodes are then seeded along these spines covering the entirety of region B. % in a pres manner.
Each spine can rotate around its fixed node on the rigid wall, while all the nodes along each spine can move with this spine as the elastic beam is deformed. This allows the mesh to adapt during deformation of the beam.

The Pertrov-Galerkin weighted residual method is used to discretise the system governing equations (\ref{FGvEq}, \ref{lBGvEq:1}-\ref{lBGvEq:2}, \ref{eq:theta} and \ref{BGvEq:thetalamkapa}). We use 6-node triangular elements with second-order shape functions for the fluid velocity components $u_1$ and $u_2$, and linear shape functions for the fluid pressure $p$. The beam variables $x_b,y_b,\theta,\lambda$ and $\kappa$ are all discretised using second-order shape functions evaluated on 3-node beam elements \citep{huyakorn1978comparison}. The weighted residuals method is used to discretise the system to determine the nodal values of all variables, where we obtain the discretised matrix equation in the form 
\begin{align}
{\mathbf M}\frac{{\rm d}{\mathbf \Theta}}{{\rm d}t}+{\mathbf K}({\mathbf \Theta})-{\mathbf F}={\mathbf R}\approx{\mathbf 0},\label{B_Num:DisMaxEq}
\end{align}
where ${\mathbf \Theta}=(u_{xi},p_{i},u_{yi},x_{bi},y_{bi},\theta_{i},\lambda_i,\kappa_{i})^{\rm{T}}$ is the global vector of unknowns with dimension $N=n_{total} \times 8$,  $i=1, ...n_{total}$ is the $i$th nodal number of the total $n_{total}$ nodes, and 8 is the degree of freedom per node;  $\mathbf M$ and $\mathbf K$  are the $N \times N$ mass and stiffness matrices, $\mathbf F$ is the external force vector and $\mathbf R$ is the residual vector.

An implicit finite difference scheme is used for time integration of the discrete nonlinear matrix equation. At each time step, a frontal method is used to assemble the global matrix equation and a Newton--Raphson iteration scheme is applied to solve the global matrix equation for $\mathbf \Theta$ \citep{luo1996numerical, hao2016arnoldi}. The numerical method enables us to evaluate the terms in the fully non-linear energy budget in (\ref{FulSysEnerBug:0}) by post-processing. As in \citet{luo2008cascade}, we use 36657 second-order 6-node triangle elements for the fluid domain and 140 second-order 3-node elements for the beam. The convergence of the numerical method is demonstrated in appendix \ref{sec:appB}.
 
\section{Results}
\label{sec: B_NumResults}

Following \citet{luo2008cascade}, throughout this study we fix the dimensionless parameters to be $L_u=L=5$, $L_d=30$, $h=0.01$, $T=0$, $c_{\kappa}=(h^2/12)c_\lambda$ and $p_e=1.95$. In what follows, we focus mainly on a small parameter region with either $c_\lambda=500$ or $c_\lambda=1600$ and $190 < Re \leq 400$ (though we also briefly summarise the parameter space spanned by $c_\lambda$ and $Re$ for fixed $p_e$). For $c_\lambda=1600$, \citet{luo2008cascade} identified an unsteady mode-2 oscillation in this region using the linear constitutive fluid-beam model of the same geometrical configuration with critical Reynolds number $Re_l\approx212.0$. We first focus on a slice through this parameter space for fixed $c_\lambda$, considering the steady (\S\ref{sec:st}) and unsteady behaviour of the system (\S\ref{sec:unst}). We then provide an overview of the regions of instability in the parameter space spanned by the extensional stiffness and Reynolds number (\S \ref{sec:overview}), summarise the dynamics of oscillations which grow from the upper branch of static solutions (\S \ref{sec:upper}), elucidate the nonlinear bifurcation structure of the system (\S \ref{sec:bifurcation}) and examine the possibility of homoclinic orbits (\S \ref{sec:homoclinic}). Finally, we summarise the associated energy budget of fully-developed oscillations (\S\ref{sec:energy}). 

\subsection{Steady solutions for $c_{\lambda}=1600$}
\label{sec:st}

\begin{figure}
\centering
\includegraphics[width=0.49\textwidth]{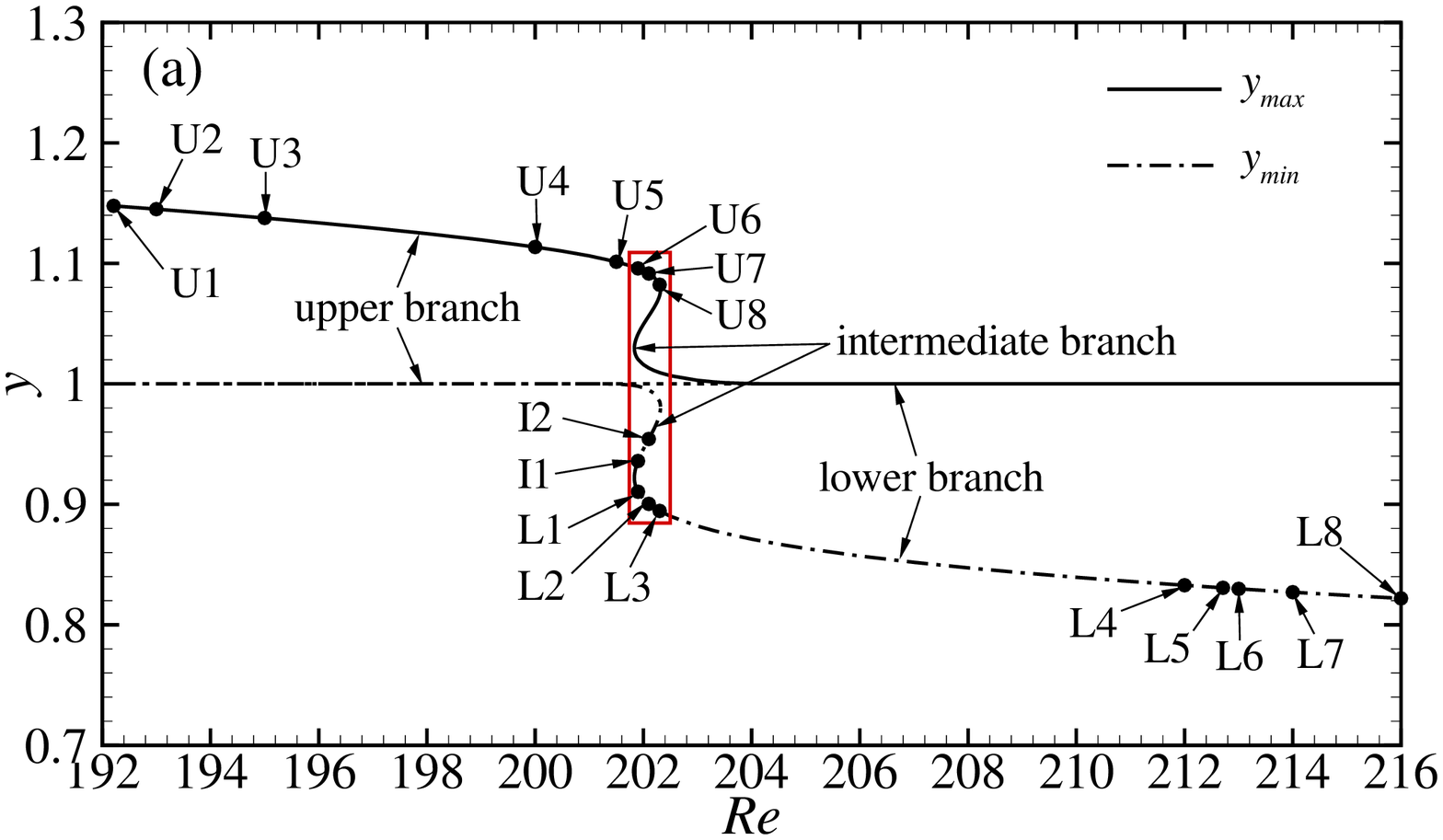}
\includegraphics[width=0.49\textwidth]{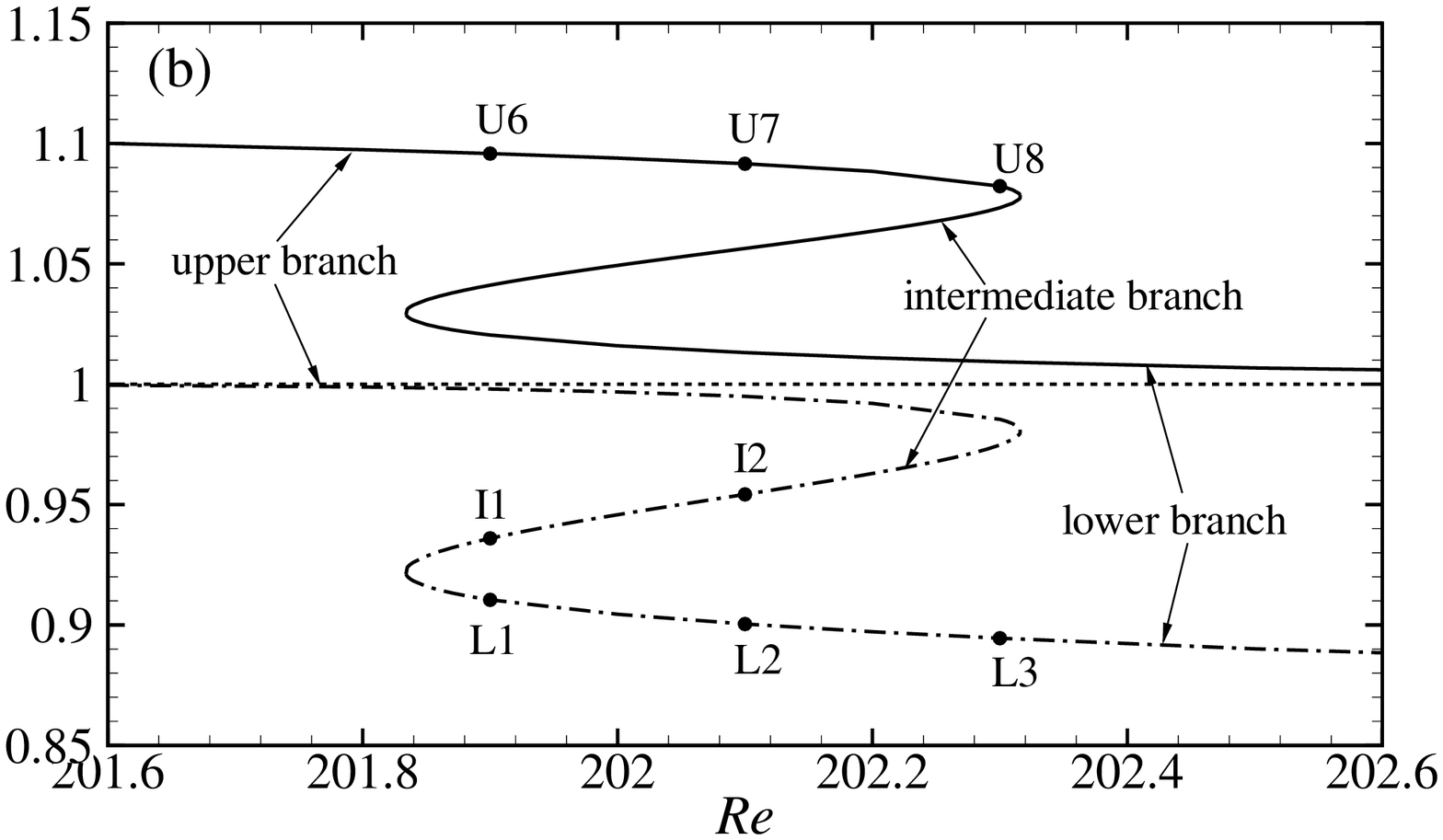}\\
\includegraphics[width=0.98\textwidth]{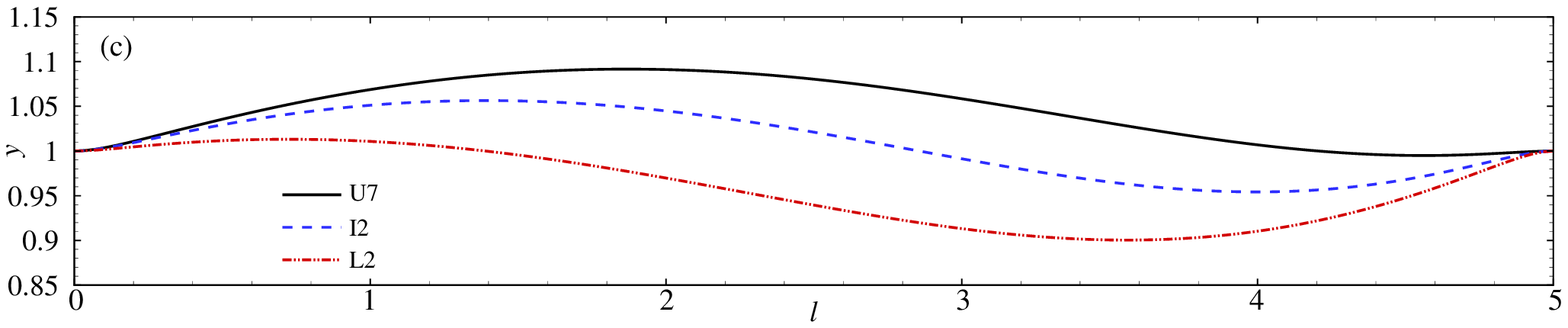}
\caption{Static solutions of the model for $c_{\lambda}=1600$ plotting: (a) the minimal ($y_{min}$) and maximal ($y_{max}$) channel widths as a function of Reynolds number, plotted with dot-dashed and solid lines, respectively, where the upper and lower static branches are labelled; (b) zoom-in of the region with multiple static solutions marked by a red square in panel (a); (c) the steady beam shape for operating points U7, I2 and L2.}
\label{Fig:st-ymin-max}
\end{figure}

%\peter{Can you rework (c) and (d) into one wide panel and show only U7, I2 and L2.}

% static solutions
In order to assess the static behaviour of the system for fixed elastic properties of the beam, figure \ref{Fig:st-ymin-max} summarises steady solutions of beam deflection with $c_{\lambda}=1600$. In particular, we consider the maximal ($y_{max}$) and minimal ($y_{min}$) steady beam position as a function of Reynolds number in figure \ref{Fig:st-ymin-max}(a), with a zoom in around the region with multiple steady solutions shown in figure \ref{Fig:st-ymin-max}(b). For low Reynolds numbers the steady beam is entirely inflated (hence $y_{min}=1$). As the Reynolds number increases the channel becomes increasingly constricted as the beam is drawn toward the rigid wall by the Bernoulli effect. For $Re\approx201.5$ the steady beam shape becomes so-called mode-2, with two extrema, which are inflated at the upstream end and collapsed at the downstream end, i.e. $y_{min} < 1$. As the Reynolds number increases further the steady solution abruptly changes at $Re\approx202.316$, transitioning to a much more collapsed configuration. This collapsed configuration persists as the Reynolds number decreases until a second transition at $Re\approx201.834$, resulting in a narrow region of parameter space with more than one steady solution. In line with \citet{stewart2017instabilities}, we term the steady solution which persists to low Reynolds numbers as the upper branch solution (where the channel wall is inflated), and the solution which persists to large Reynolds numbers as the lower branch solution (where the channel wall is collapsed). These two transition points take the form of limit point bifurcations and are termed the upper and lower limit points, respectively. The upper and lower branches are connected by an intermediate branch. A three branch static structure has previously been reported for the collapsible channel system, both using a full two-dimensional fluid models with simplified wall models \citep{luo2000multiple,heil2004efficient}, as well as using a reduced one-dimensional model for the flow \citep{stewart2010flows,stewart2017instabilities}. Such three branch behaviour was also recently demonstrated by \cite{herrada2021global} using a hyperelastic (neo-Hookean) wall of finite thickness. In summary, this figure demonstrates that the steady system exhibits at least one static solution across the parameter space, with a narrow region where it can exhibit three static configurations.

% steady points
To illustrate the behaviour of the system for $c_\lambda=1600$ we select eight points on the upper branch of static solutions, which we term U1-U8, two points on the intermediate branch, which we term I1 and I2, and eight points on the lower branch of static solutions, which we term L1-L8. These points are labelled on figure \ref{Fig:st-ymin-max} and their corresponding values of Reynolds number are listed in table \ref{Tab:energy} below. 

% steady wall shape
To assess these static configurations in detail, figure \ref{Fig:st-ymin-max}(c) illustrates the three possible steady beam shapes for $Re=202.1$, $c_\lambda=1600$ (operating points U7, I2 and L2). On the each branch of steady solutions the wall shape is so-called mode-2, bulged out near the upstream and collapsed at the downstream end of the channel. The upper branch is more inflated than the intermediate branch, which is itself more inflated than the lower branch. Similarly, the lower branch is more collapsed than the intermediate branch, which is in turn more collapsed than the upper branch.

We will analyse the stability and the energy budget of these branches in later sections and the region of parameter space with more than one static solution is shown in figure \ref{Fig:st-saddle-neutral} below.

\subsection{Unsteady solutions for $c_{\lambda}=1600$}
\label{sec:unst}

\begin{figure}
\centering
\includegraphics[width=0.98\textwidth]{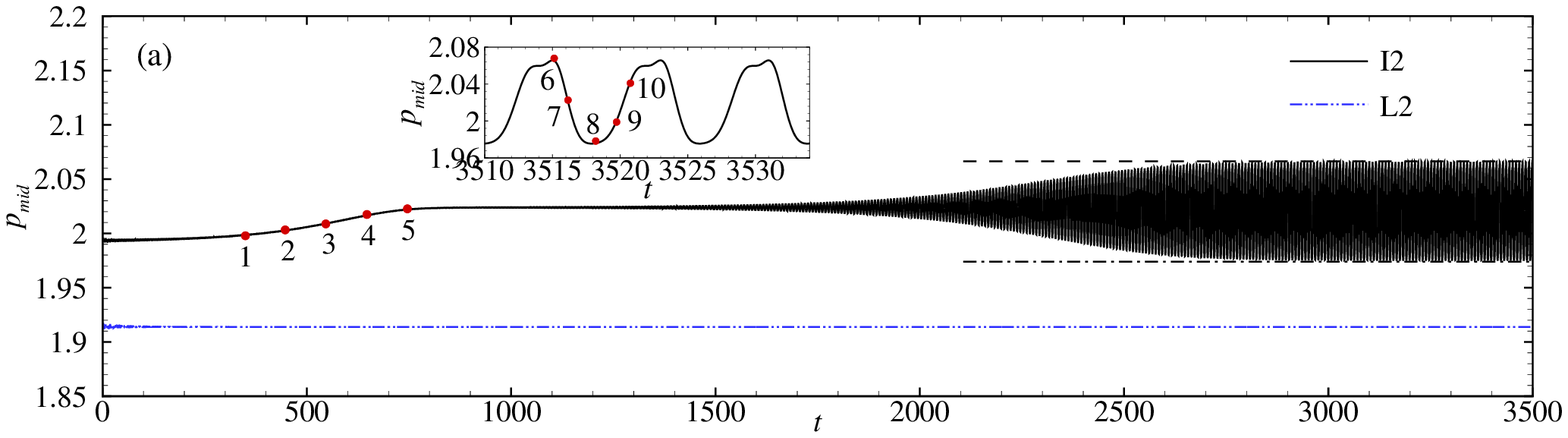}\\
\includegraphics[width=0.47\textwidth]{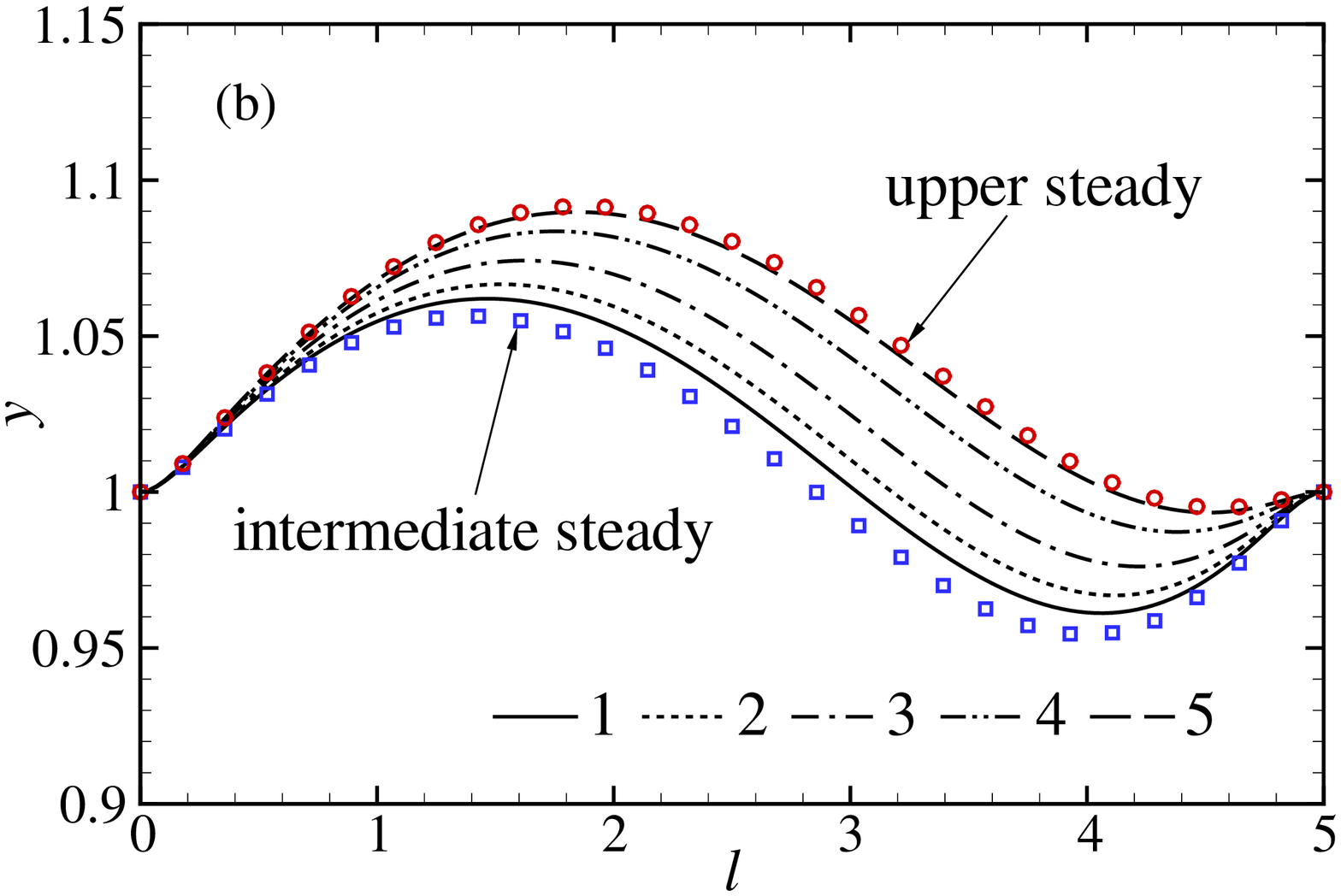}
\includegraphics[width=0.47\textwidth]{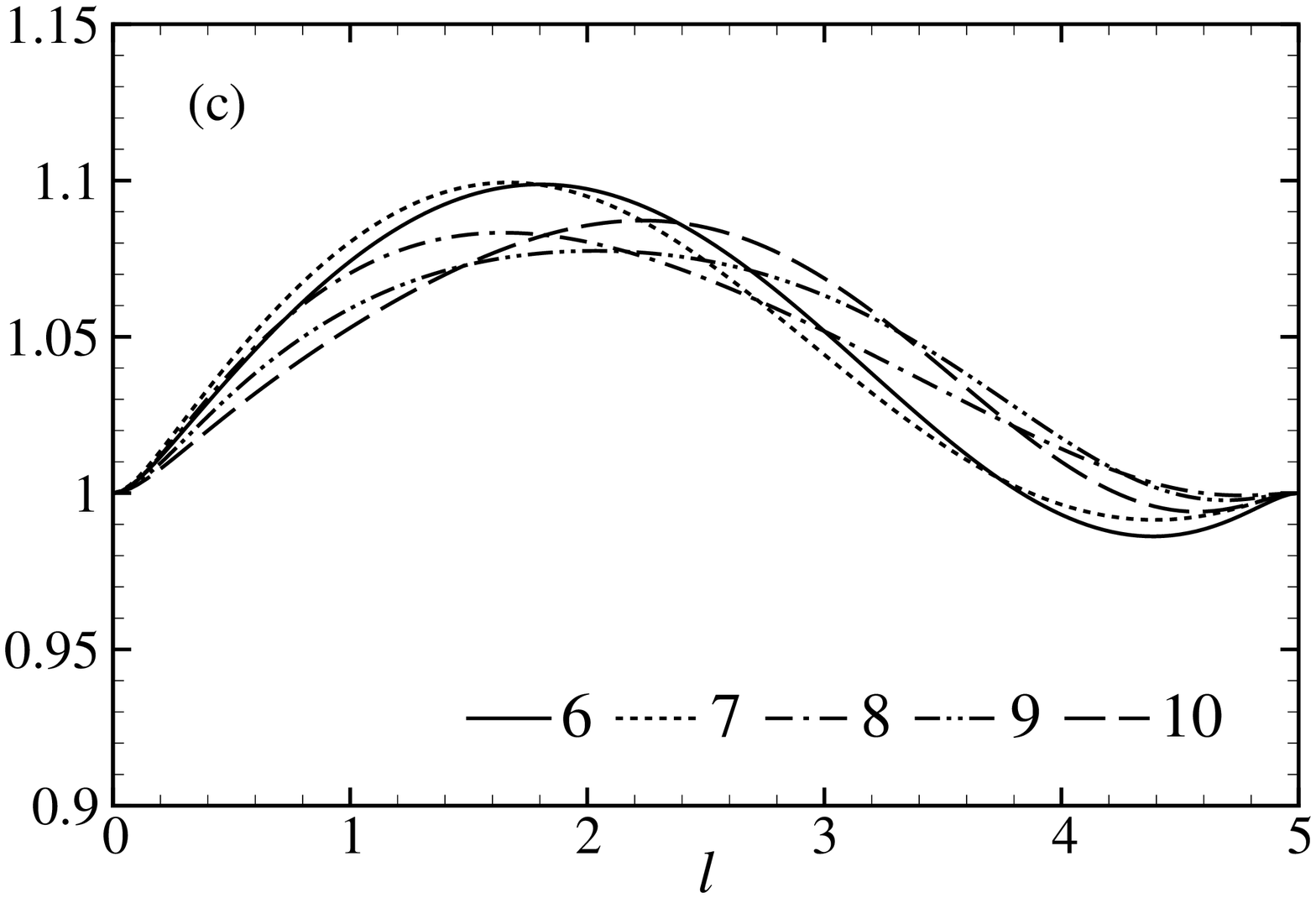}
\caption{Unsteady solutions at $Re=202.1$, $c_{\lambda}=1600$ (operating points U7, I2 and L2) showing: (a) time-trace of the wall mid-point pressure $p_{mid}$ initiated close to the intermediate static branch point I2; (b) five wall profiles as the system evolves from the intermediate static branch toward the upper branch static state; (c) . The dashed (dot-dashed) lines in (a) show the maximal (minimal) midpoint pressure from the upper branch limit cycle U7. The beam profile plotted with open squares (circles) in (b) show the corresponding intermediate (upper) static configurations.}
\label{fig:intermediate}
\end{figure}

% setup
In order to test the stability of the system to time-dependent perturbations, we apply a small increment to the steady solution to generate an initial condition for the computations (here we use the steady solution along the same branch with a $1\%$ increase in $c_{\lambda}$). As is conventional in hydrodynamic stability theory, the system is deemed stable if the unsteady solution converges to the corresponding steady solution following the perturbation, and unstable if the perturbation grows with time \cite[][]{drazin2002introduction}.  The boundary between these two behaviours is termed neutrally stable. In plotting the unsteady behaviour of the system we generally illustrate only the fully developed limit cycle of the oscillations, truncating the period of transient growth from the initial condition. Exceptions to this are given in figures \ref{fig:intermediate} and \ref{Fig:mesh-validation}(b) below, where we show the full dynamics from the imposed initial condition.

\begin{figure}
\centering
\includegraphics[width=0.325\textwidth]{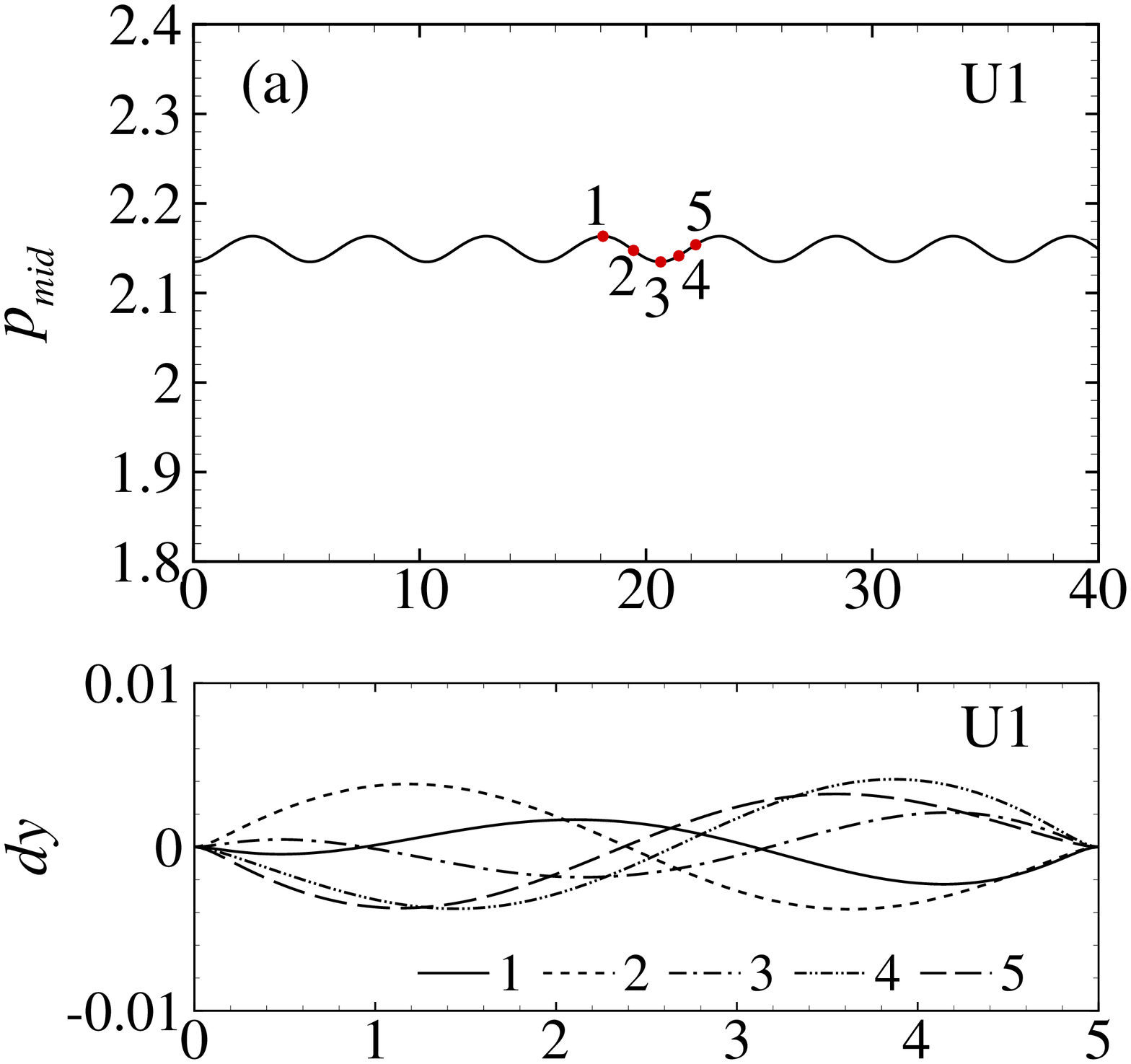}
\includegraphics[width=0.325\textwidth]{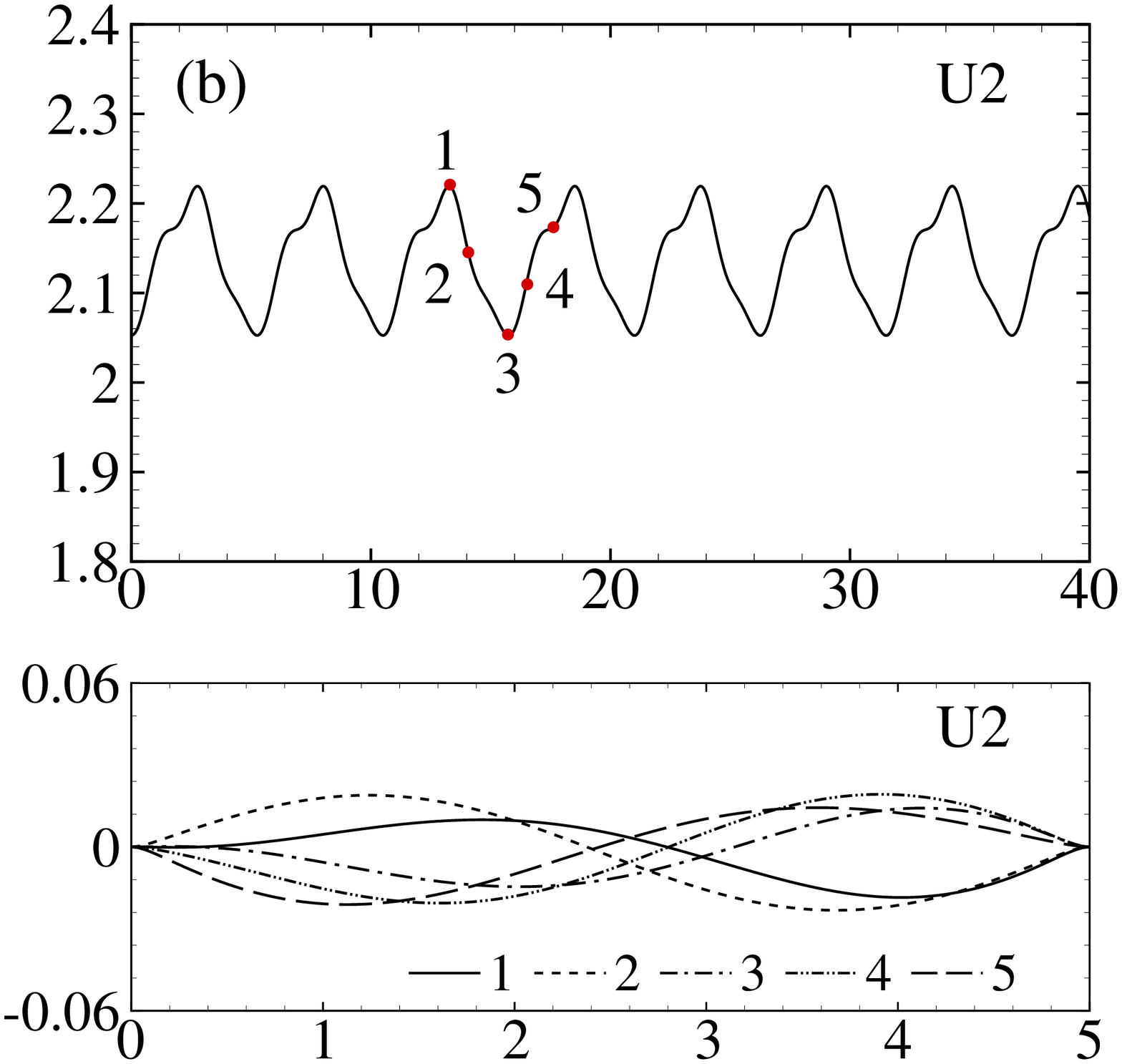}
\includegraphics[width=0.325\textwidth]{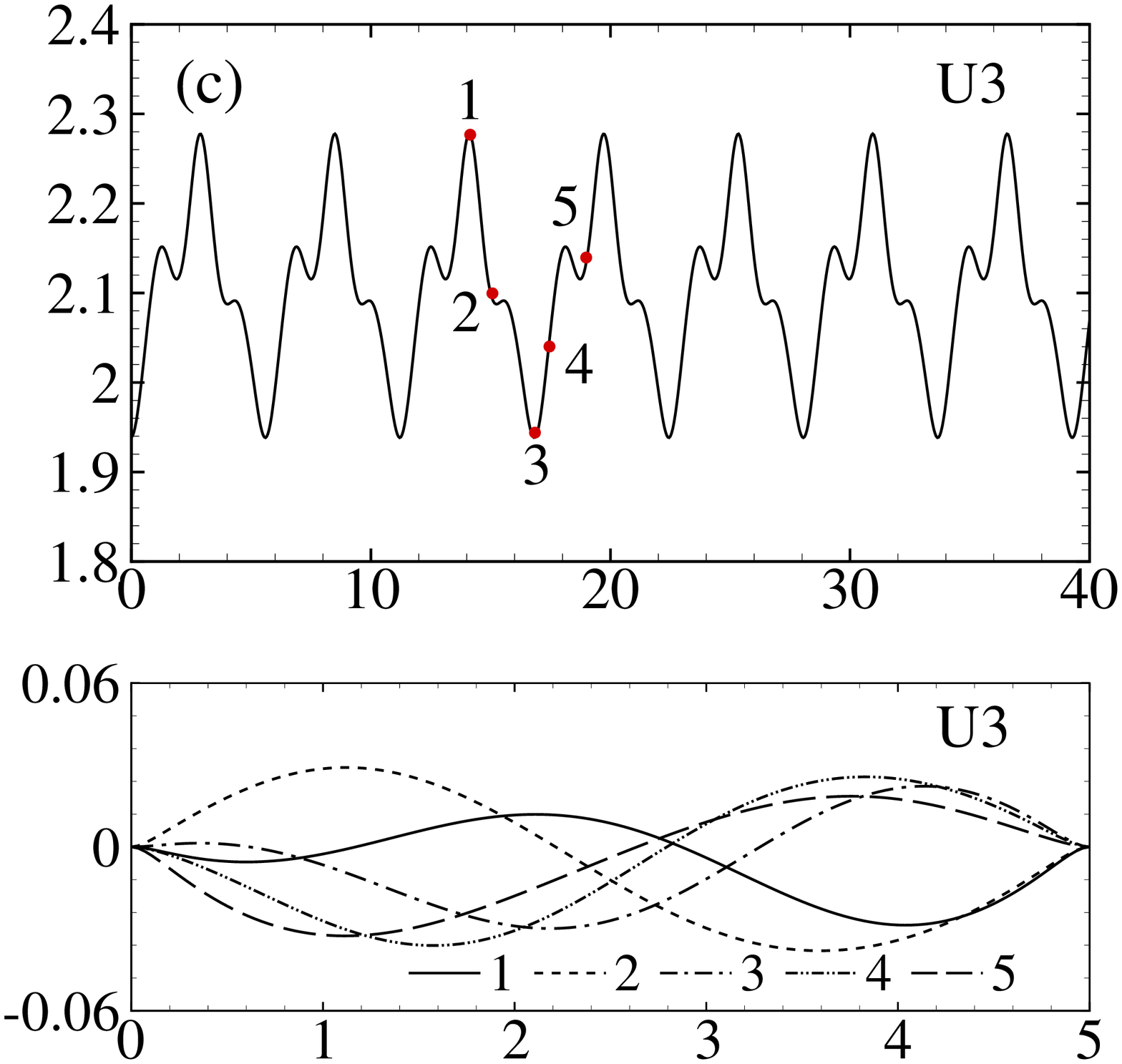}\\
\includegraphics[width=0.325\textwidth]{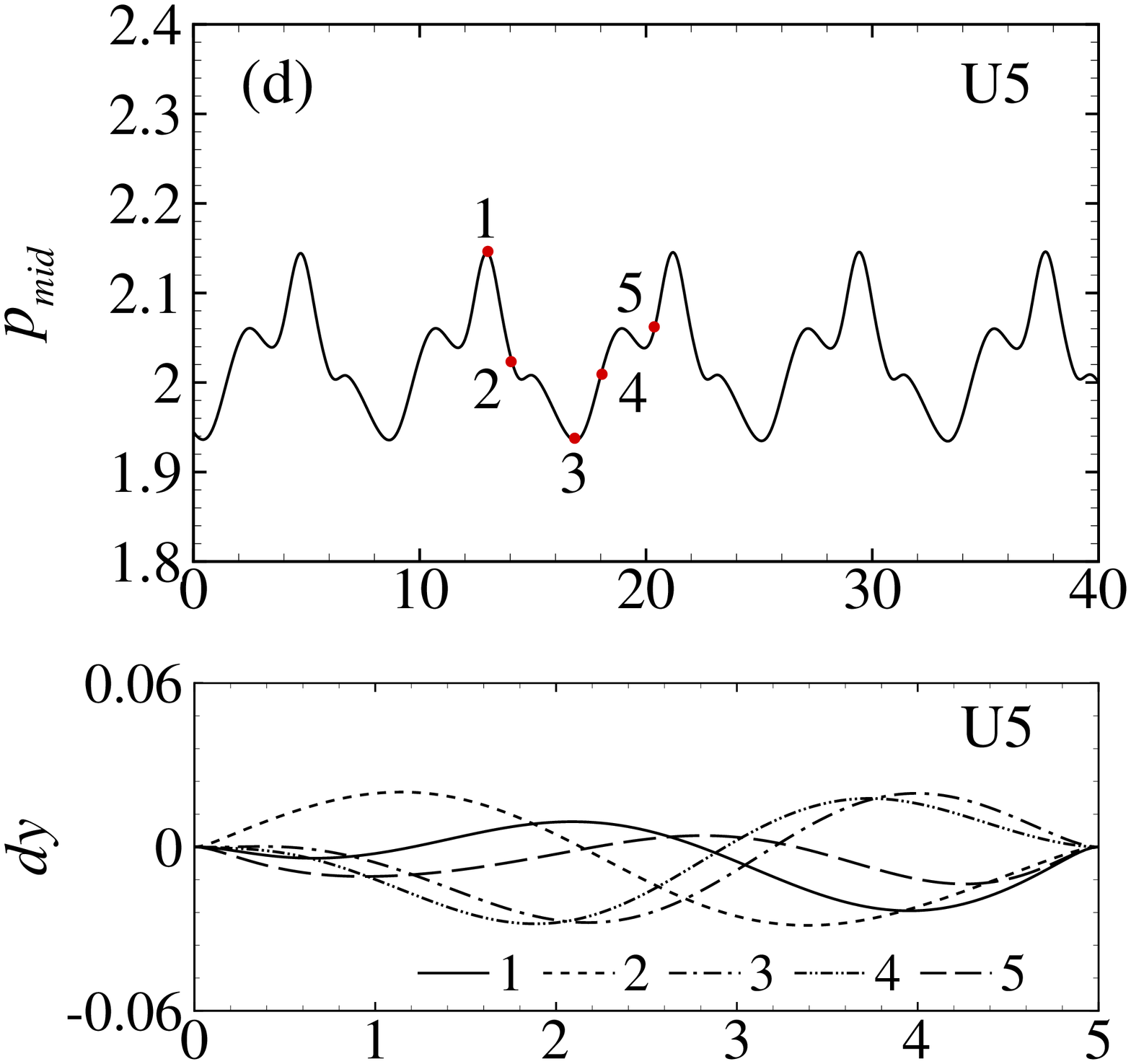}
\includegraphics[width=0.325\textwidth]{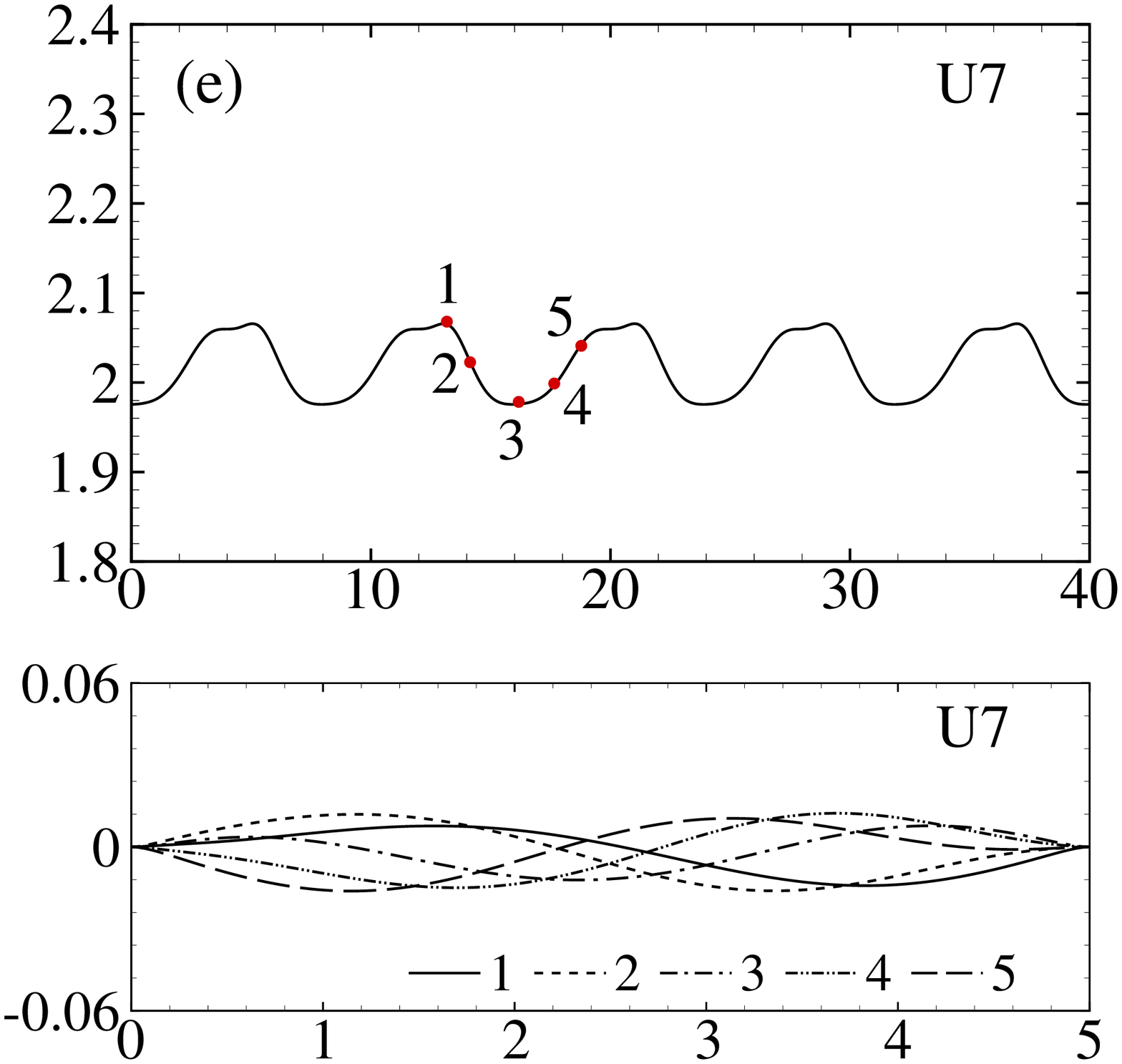}
\includegraphics[width=0.325\textwidth]{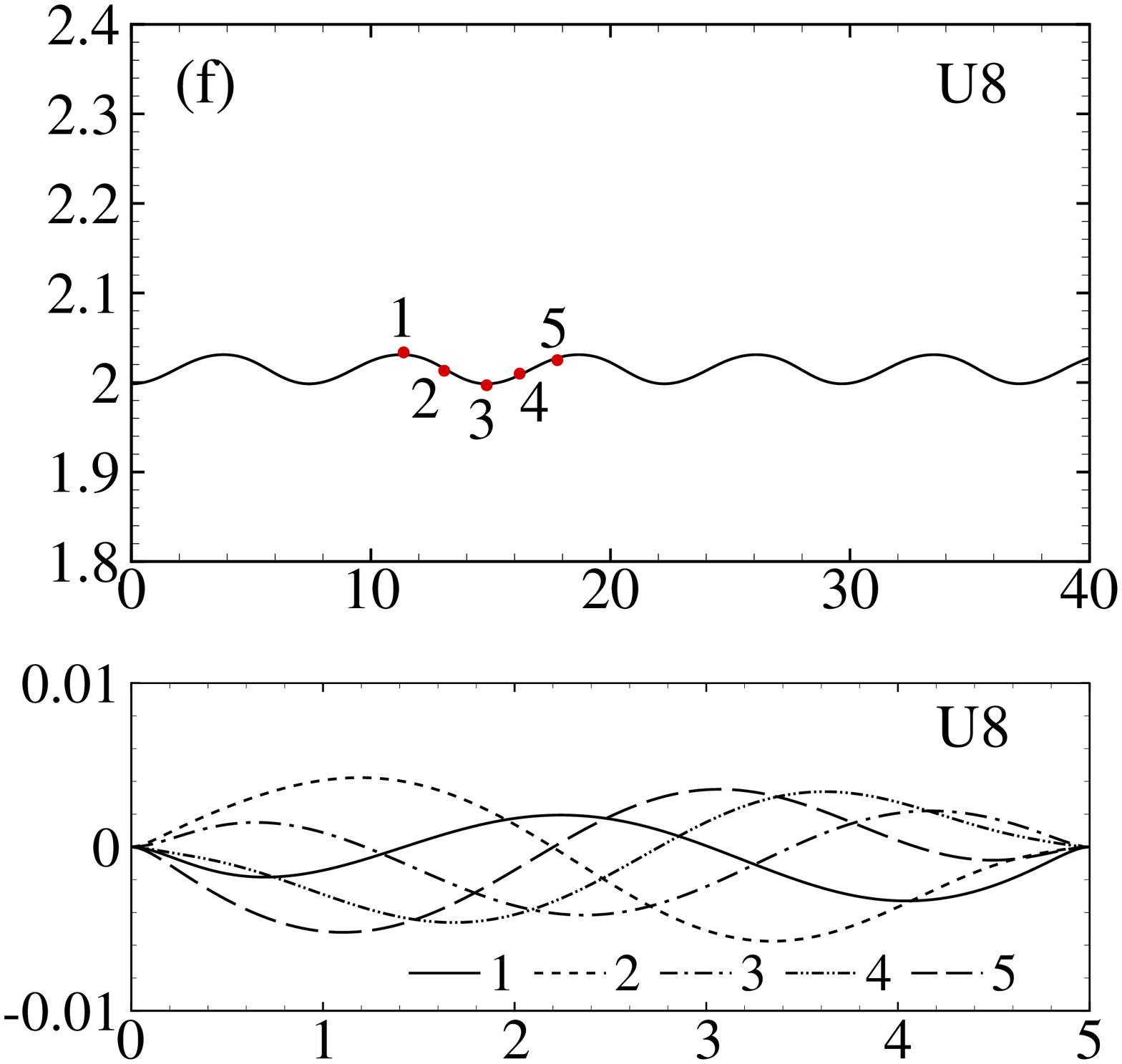}\\
\includegraphics[width=0.325\textwidth]{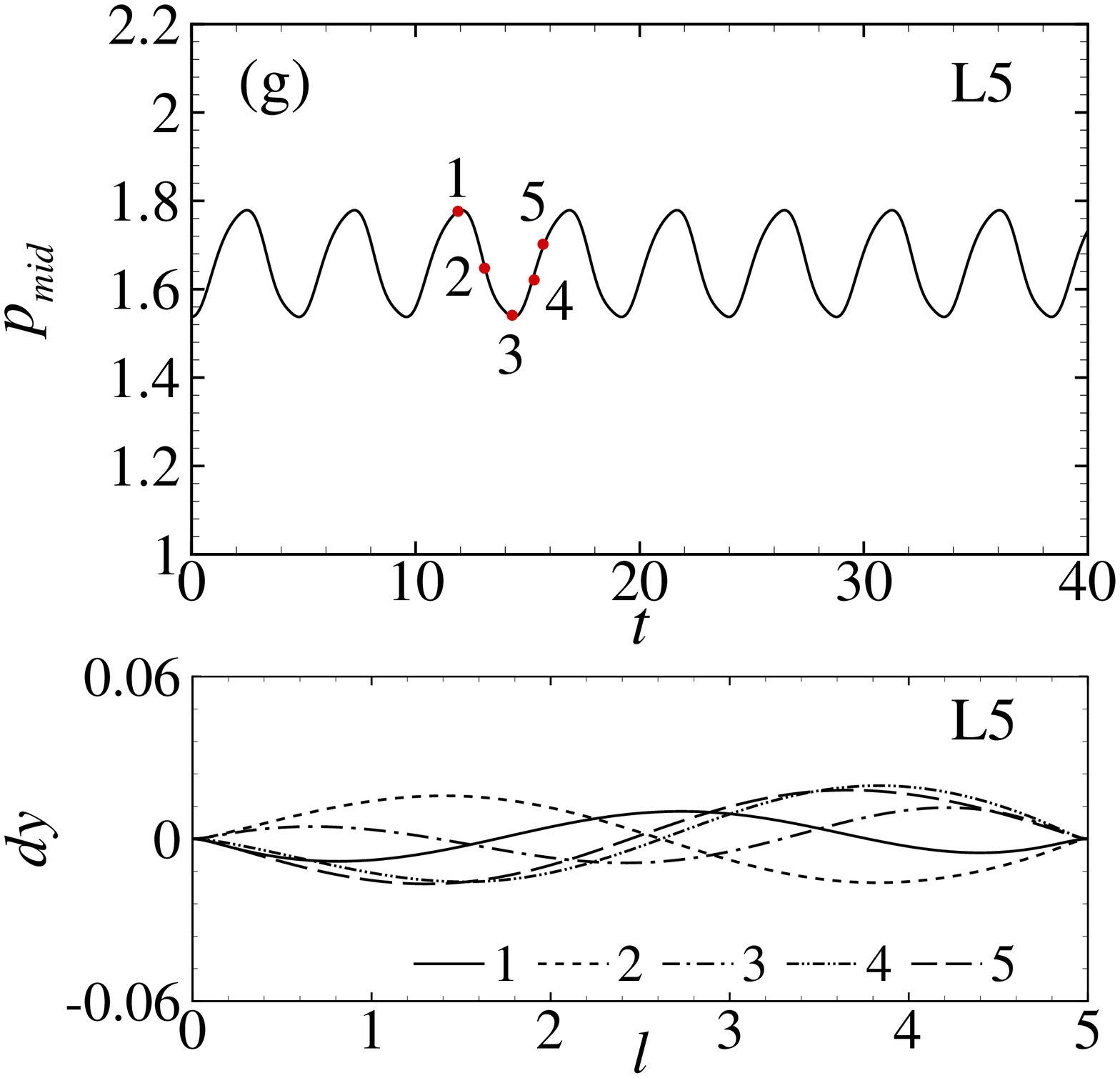}
\includegraphics[width=0.325\textwidth]{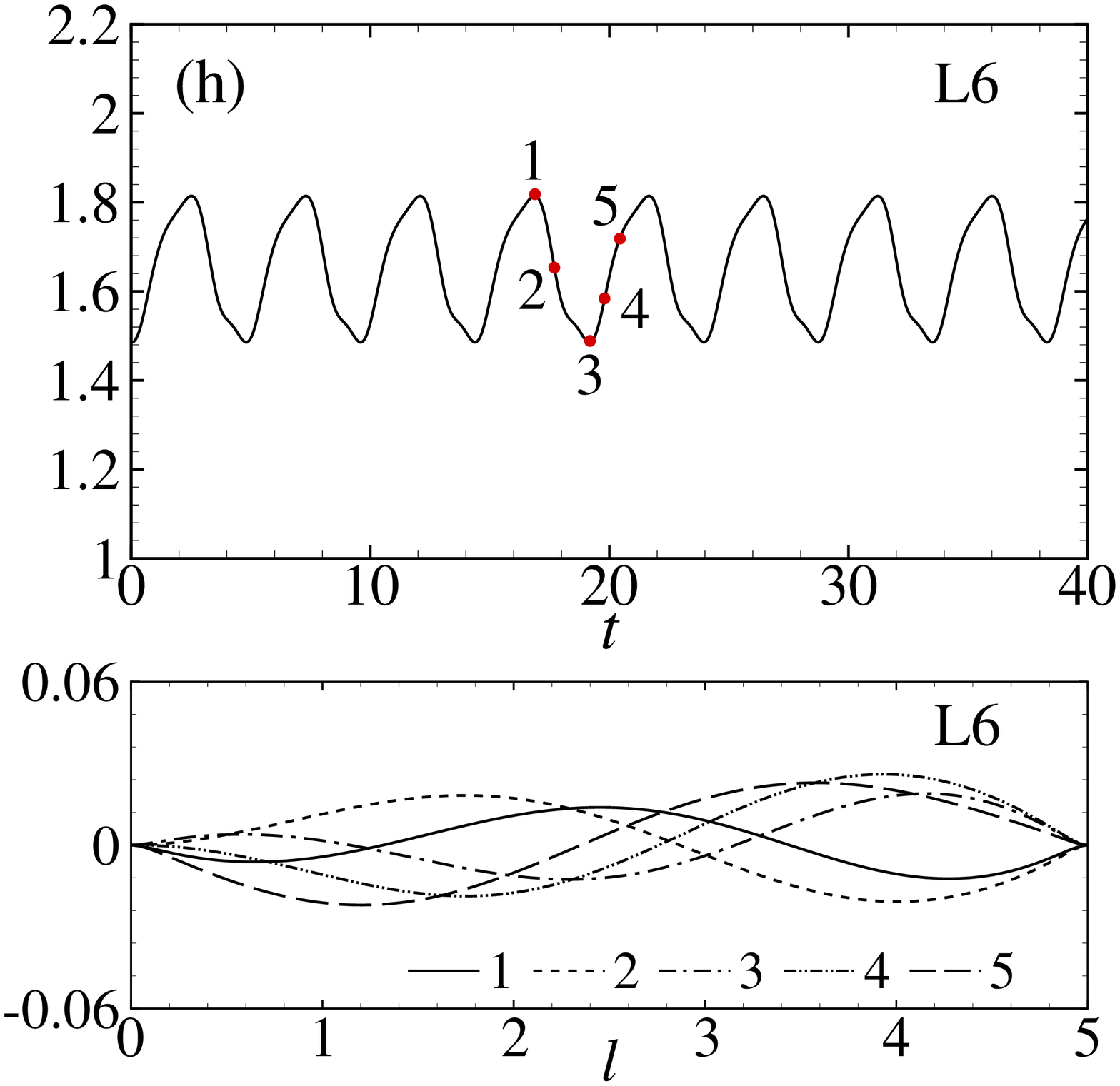}
\includegraphics[width=0.325\textwidth]{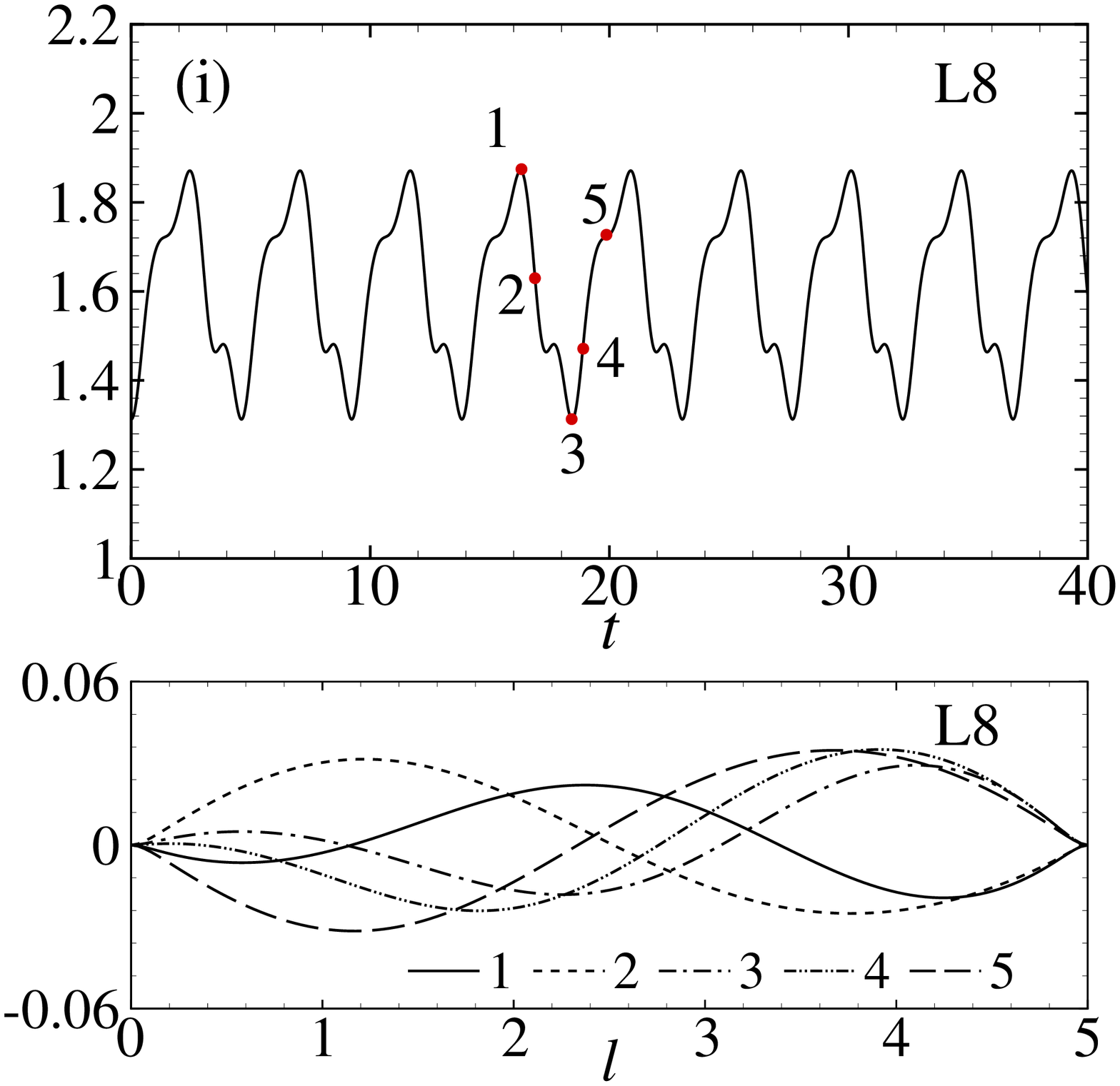}
\caption{Dynamics of self-excited oscillations arising from the upper and lower static branches for $c_{\lambda}=1600$, showing time-traces of the wall mid-point pressure (upper plot for each sub-panel) and the corresponding perturbation wall shape $dy = y_b-y_b^{(0)}$ at five selected time instances labelled 1-5 (lower plot in each sub-panel) for operating points: (a) U1, $Re=192.21$; (b) U2, $Re=193$; (c) U3, $Re=195$; (d) U5, $Re=201.5$; (e) U7, $Re=202.1$; (f) U8, $Re=202.3$; (g) L5, $Re=212.71$; (h) L6, $Re=213$; (i) L8, $Re=216$.}
\label{Fig:pmid-t}
\end{figure}

% stability of upper and lower branches
In order to assess the stability of the upper and lower branches of static solution (identified in figure \ref{Fig:st-ymin-max}), in figure \ref{Fig:pmid-t} we plot time-traces of the fluid pressure on the elastic wall at the midpoint of the beam ($x=L/2$) for various values of the Reynolds number with $c_\lambda=1600$. We further illustrate various unsteady perturbation wall profiles over the period of fully developed oscillation by subtracting the corresponding steady wall profile, i.e. $dy=y_b-y_b^{(0)}$. For sufficiently low Reynolds numbers the upper branch static solutions are stable (not shown).  The steady configuration on the upper branch becomes unstable at $Re\approx192.2$ (figure \ref{Fig:pmid-t}a), just outside the region with multiple static states. This oscillation grows in amplitude as the Reynolds number increases (figure \ref{Fig:pmid-t}b), becoming increasingly non-sinusoidal/irregular (figure \ref{Fig:pmid-t}c). Above the critical value of Reynolds number the oscillatory wall profile is mode-2 (two extrema across the compliant segment, shown below the corresponding time-traces in figures \ref{Fig:pmid-t}a-c), although the beam profile is mode-3 as it moves through the static configuration ({\it eg} profile labelled 1 in figure \ref{Fig:pmid-t}a). Proceeding along the upper branch, the amplitude of oscillation reaches a maximum at $Re \approx 199.0$ (see profiles below) and then decreases again (figures \ref{Fig:pmid-t}d-f). The unstable branch eventually enters the region with multiple static solutions and approaches zero amplitude and re-stabilises as the steady branch becomes close to the upper branch limit point ($Re\approx202.316$). The wall profiles in this region are again mostly mode-2 (although some of the profiles are mode-3), shown below the corresponding time-traces (figures \ref{Fig:pmid-t}d-f). We explore the interaction between the upper branch limit cycles and the upper branch limit point in \S \ref{sec:homoclinic} below. Conversely, the lower branch of static solutions is stable to oscillations as it emerges from the lower limit point ($Re\approx201.834$). As the Reynolds number increases, the lower branch steady solution eventually becomes unstable at $Re\approx212.65$, outside the region with multiple steady states; the oscillation profile increases in amplitude as the Reynolds number increases (figures \ref{Fig:pmid-t}h,i) and is again mostly mode-2 (see profiles below the corresponding time-traces in figures \ref{Fig:pmid-t}g-i, though the wall profile can be mode-3 as it moves through the static configuration). In summary, this figure demonstrates that both the upper and lower static branches can become unstable to oscillations in the neighbourhood of (but just outside) the region of parameter space with multiple static solutions. 

% stability of the intermediate branch

In order to test the stability of the intermediate static branch (identified in figure \ref{Fig:st-ymin-max}), in figure \ref{fig:intermediate}(a) we plot time-traces of the fluid pressure on the elastic wall at the mid-point of the beam ($x=L/2$) initiated at operating point I2. Initially the mid-point pressure increases toward the upper static branch, where the corresponding beam profile gradually expands toward the upper branch static state (figure \ref{fig:intermediate}b). Since the upper branch of static solutions is unstable for these parameters (see figure \ref{Fig:pmid-t}), the system evolves toward the upper branch oscillatory limit cycle; five snapshots of the beam over an oscillation are shown in figure \ref{fig:intermediate}(c), analogous to the beam profiles identified for the upper branch limit cycle identified around operating point U2 (figure \ref{Fig:pmid-t}b). It emerges that for this model the intermediate branch is always unstable for all the parameters tested, consistent with earlier predictions in flow through flexible-walled channels \cite[][]{stewart2017instabilities,herrada2021global}, with the profile evolving to the upper branch limit cycle. 

\subsection{Overview of the parameter space}
\label{sec:overview}

\begin{figure}
\centering
\includegraphics[width=0.95\textwidth]{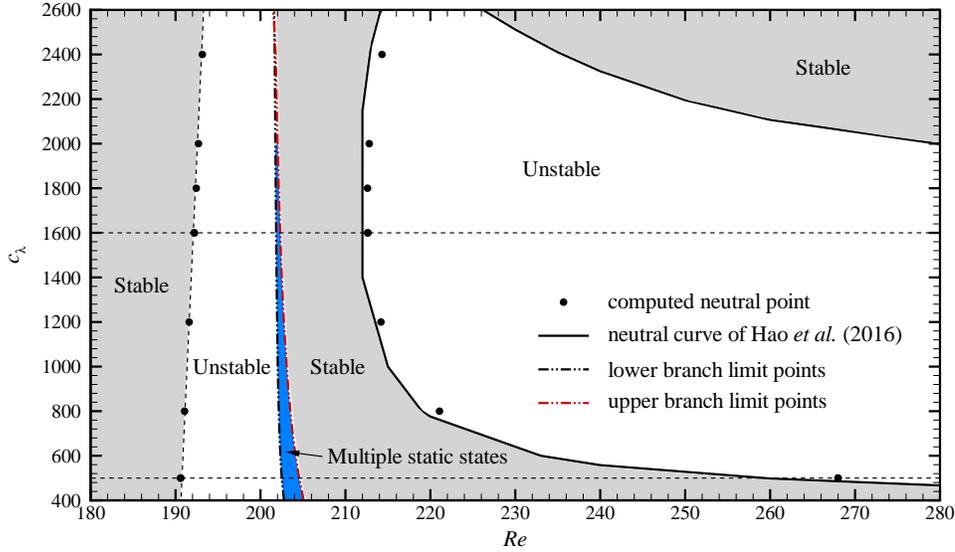}
\caption{An overview of the parameter space spanned by the Reynolds number and extensional stiffness, summarising the steady and unsteady solutions of the model. The limit points of the upper and lower static branches are plotted as red and black dot-dashed lines, respectively, and the region with multiple static solutions is shaded in light blue. The neutral stability curve initially identified by \citet{luo2008cascade} and refined by \citet{hao2016arnoldi} is plotted as a solid line (associated with the lower static branch). The computed neutral points from the current model are marked as filled black circles. The neutral stability curve associated with the upper static branch is estimated as the dotted line between the computed neutral points. The regions where the system is stable to self-excited oscillations are shaded in grey.}
\label{Fig:st-saddle-neutral}
\end{figure}

% overview of the parameter space, lower branch
Following \cite{luo2008cascade}, in figure \ref{Fig:st-saddle-neutral} we overview the stability of the system in the parameter space spanned by the Reynolds number ($Re$) and extensional stiffness ($c_\lambda$), where they showed that the space could be partitioned into a number of unstable tongues. These predictions were updated slightly by \cite{hao2016arnoldi} using a global stability eigensolver, and their neutral stability curve is shown as a solid black line in figure \ref{Fig:st-saddle-neutral}. The corresponding neutral stability points from the lower static branch computed using our numerical method are shown as filled black circles. These points agree well with the neutral stability curve of \cite{hao2016arnoldi}, and the slight differences are attributed to the difference between the two constitutive laws (the critical Reynolds number is displaced by less than 1\% for all points tested). In general the lower static branch becomes unstable as the Reynolds number increases \cite[similar to][]{heil2004efficient,stewart2017instabilities,herrada2021global}. However, \cite{luo2008cascade} \cite[see also][]{hao2016arnoldi} showed that this neutral stability curve is non-monotonic and for large $c_\lambda$ the system restabilises again as the Reynolds number becomes sufficiently large, though we did not investigate this regime.

% upper branch
However, as noted in figure \ref{Fig:pmid-t}, the upper branch of static solutions can also become unstable to oscillations. The corresponding computed neutral stability points for the upper static branch are plotted on figure \ref{Fig:st-saddle-neutral}, revealing a new region of instability to the left of the region noted by \cite{luo2008cascade} and \cite{hao2016arnoldi}.  

% extend understanding of parameter space, multiple static solutions
To extend our understanding of this parameter space in figure \ref{Fig:st-saddle-neutral} we also highlight the region of parameter space with more than one static solution, by tracing the upper and lower branch limit points as a function of the Reynolds number and the extensional stiffness $c_\lambda$ for all other parameters held fixed. It emerges that the region which admits multiple static solutions (shaded in figure \ref{Fig:st-saddle-neutral}) is very narrow. As $c_\lambda$ increases the upper and lower limit points approach each other and for $c_\lambda \gtrsim 3500$ the static solution becomes unique for all Reynolds numbers (not shown). Conversely, the width of the region with multiple static states expands slightly as $c_\lambda$ decreases, but is confined to $202.495\leq Re \leq 204.554$ for $c_\lambda=500$. 

\subsection{Self-excited oscillations of the upper branch of static solutions}
\label{sec:upper}

\begin{figure}
\centering
\includegraphics[width=0.98\textwidth]{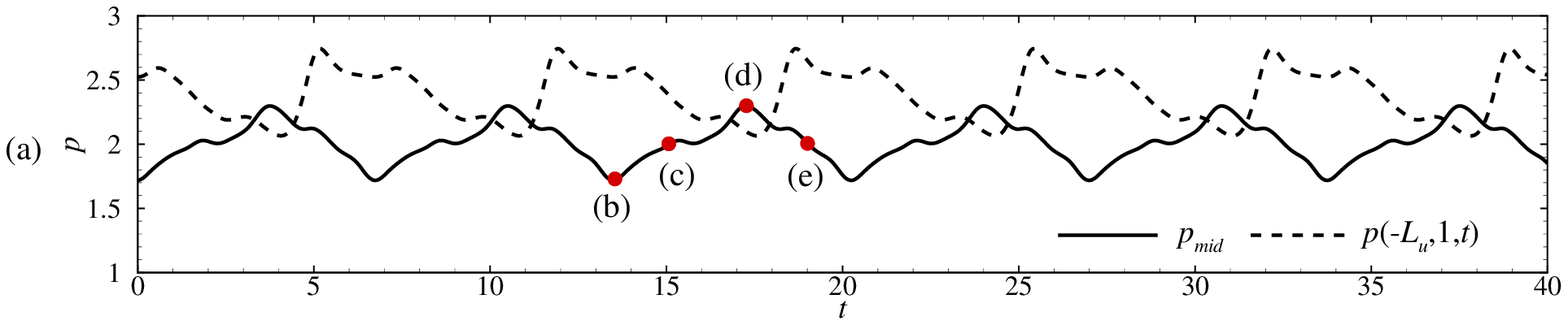}\\
\includegraphics[width=0.98\textwidth]{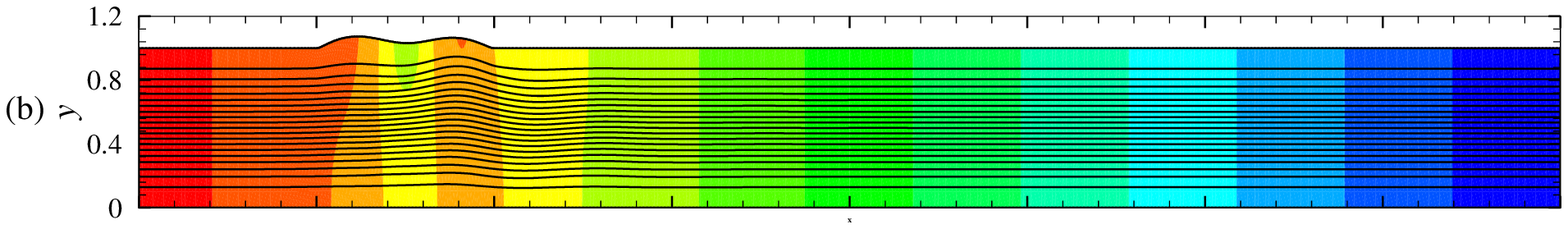}\\
\includegraphics[width=0.98\textwidth]{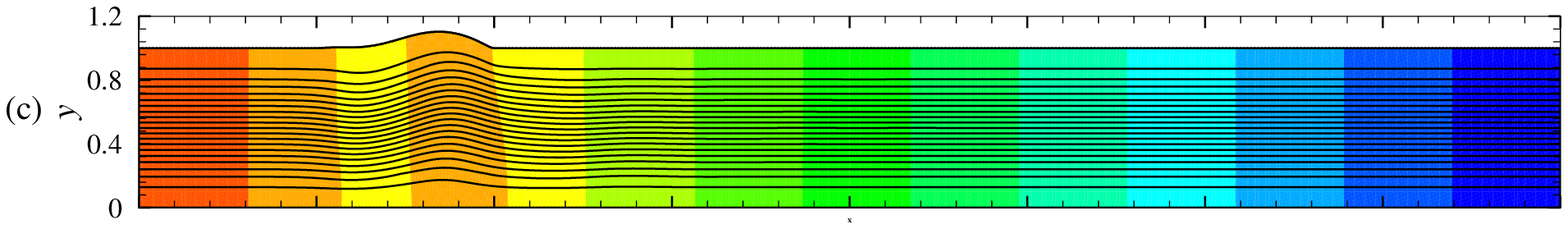}\\
\includegraphics[width=0.98\textwidth]{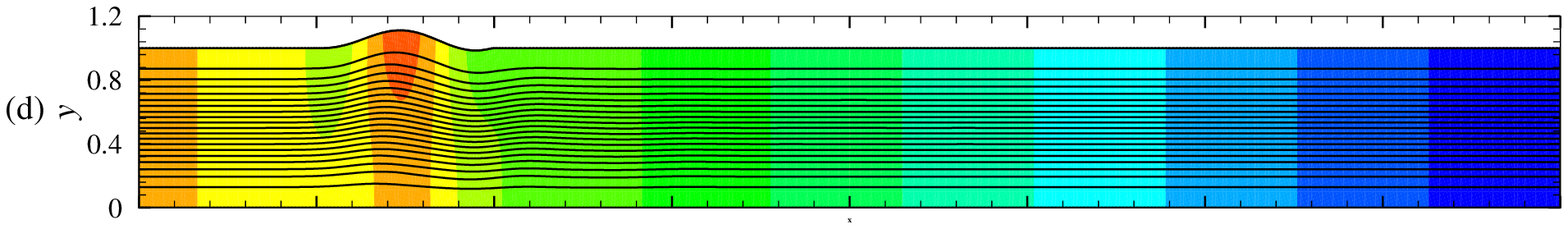}\\	
\includegraphics[width=0.98\textwidth]{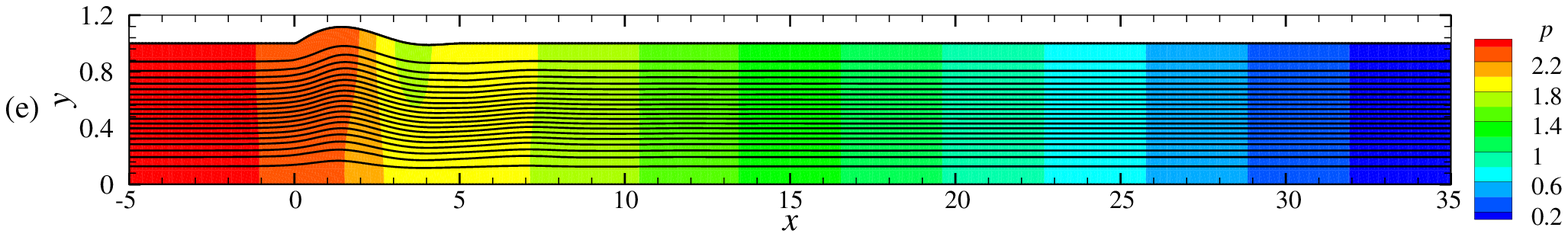}
\caption{Dynamics of self-excited oscillations from the upper static branch with $Re=199$ for $c_{\lambda}=1600$, plotting (a) the temporal evolution of mid-point wall pressure (solid line) and the upstream driving pressure evaluated on the upper wall (dashed line); (b)-(e) streamlines and pressure contours of the flow at four equally-spaced time instances over a period of oscillation, labelled panel in (a).}
\label{fig:upper}
\end{figure}

The flow-fields and pressure contours associated with oscillations of the lower branch of static solutions have been presented elsewhere \citep{heil2004efficient,luo2008cascade}. Self-excited oscillations originating from the upper branch of static solutions are explored in more detail in figure \ref{fig:upper}, where we show time-traces of the channel inlet pressure on the wall at $y=1$ and the mid-point pressure on the flexible wall for $Re=199$ (figure \ref{fig:upper}a) as well as streamlines and pressure contours at four equally spaced times over a period of oscillation (figures \ref{fig:upper}b-e). When the mid-point pressure takes its minimal value, the flexible wall is entirely inflated but with two prominent outward bulges (figure \ref{fig:upper}b), located close to the upstream and downstream ends of the compliant segment, respectively. Each bulge is associated with a local increase in fluid pressure (with a region of lower pressure between). As time progresses the downstream bulge grows at the expense of the upstream bulge (figure \ref{fig:upper}c), translating upstream and leading to a small indentation at the downstream end of the beam, with an accompanying change in the fluid pressure profile (figure \ref{fig:upper}d). As the bulge propagates upstream the driving pressure must increase abruptly to maintain the prescribed flow (between the points marked (d) and (e) in figure \ref{fig:upper}a). However, the upstream bulge is eventually suppressed by interaction with the upstream rigid segment (figures \ref{fig:upper}e). As the upstream bulge diminishes the resulting downstream fluid motion drives a new bulge at the downstream end of the compliant segment and the process repeats. Hence, the $x$-position of the maximal channel width does not change smoothly over time (two local maxima interchange as the global maximum). The associated flow-fields are very laminar with almost no cross-stream pressure gradient (figures \ref{fig:upper}b-e). In particular, there are no regions of flow separation or vorticity waves in the downstream rigid segment, which are typically associated with lower branch flow-fields (see figure \ref{Fig:U10-nkl-okl} below). A movie of the dynamics of these upper branch oscillations is provided in online supplementary material. In summary, this figure shows that the upper branch instability involves upstream propagation of a bulge in the wall profile (which develops at the downstream end). However, this bulge does not propagate back downstream again over a period; instead it is washed out by interaction with flow through the upstream rigid segment and replaced by an entirely new bulge at the downstream end of the compliant segment. 

% bifurcation diagram
\subsection{Bifurcation structure of the dynamical system}
\label{sec:bifurcation}

\begin{figure}
\centering
\includegraphics[width=0.49\textwidth]{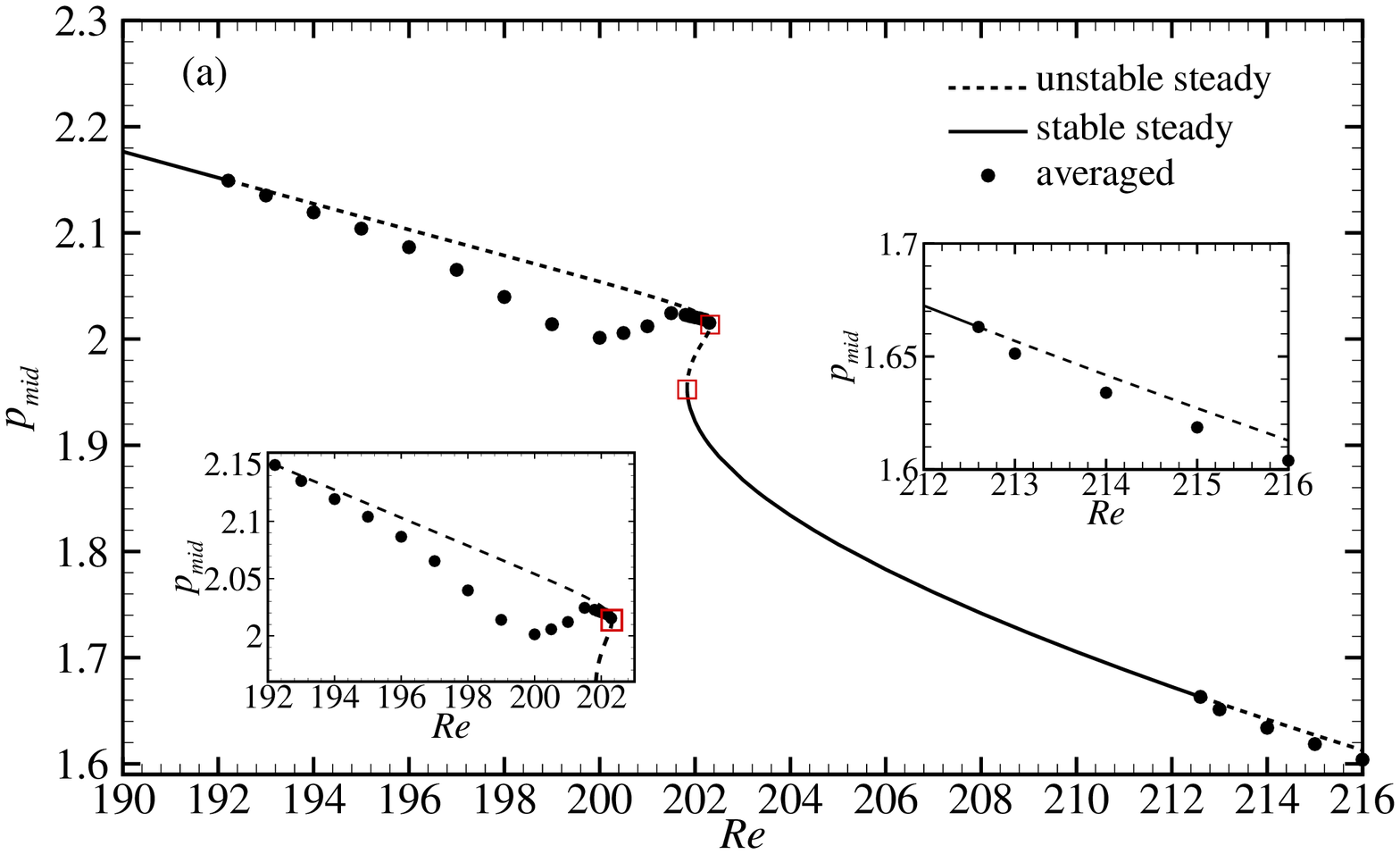}
\includegraphics[width=0.49\textwidth]{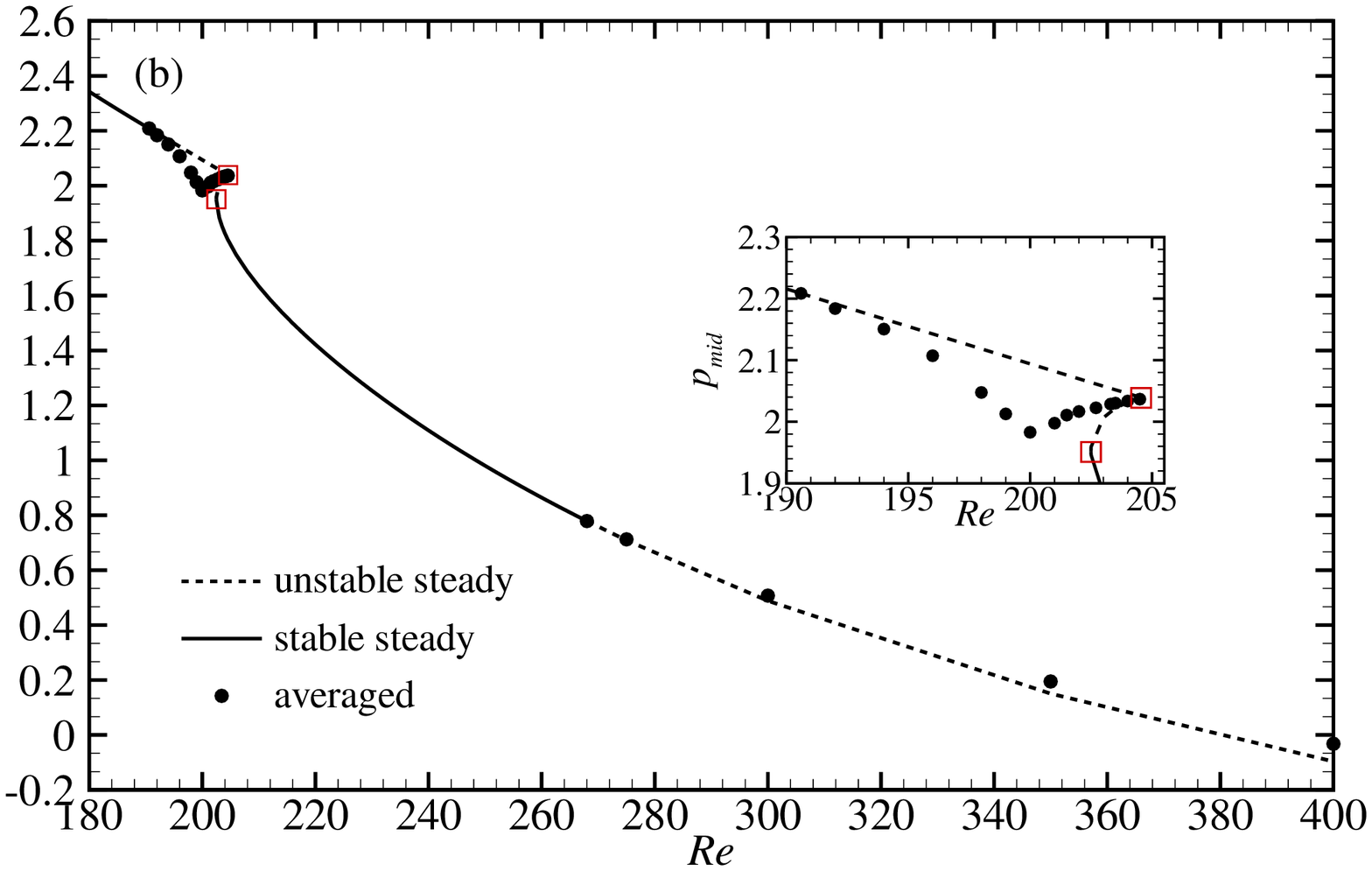}
\caption{Nonlinear bifurcation diagram plotted as a function of Reynolds number, showing the time-averaged midpoint wall pressure $p_{mid}^{(avg)}$ (filled black circles) and the static midpoint pressure $p_{mid}$ as solid (stable) and dashed lines (unstable) for: (a) $c_\lambda=1600$ and (b) $c_\lambda=500$. The upper and lower branch limit points are shown as red open squares. The insets show a close-up around the upper and lower limit points.}
\label{fig:bifurcation}
\end{figure}

To assess the relative growth of the oscillatory limit cycles, in figure \ref{fig:bifurcation} we employ the Reynolds number as a bifurcation parameter and plot the time-averaged midpoint pressure on the flexible wall for $c_\lambda=1600$ (figure \ref{fig:bifurcation}a) and $c_\lambda=500$ (figure \ref{fig:bifurcation}b) alongside the corresponding static branches of the system (a different visualisation of figure \ref{Fig:st-ymin-max}). As the Reynolds number increases along the upper static branch we observe supercritical growth of the oscillation from the neutral stability point, where the time-averaged midpoint pressure is decreased compared to the corresponding static value (figures \ref{fig:bifurcation}a,b). Proceeding along the upper branch, the oscillation amplitude approaches zero as the Reynolds number approaches the upper branch limit point (figures \ref{fig:bifurcation}a,b); the dynamics of this interaction is explored in \S \ref{sec:homoclinic} below. 

Almost identical behaviour was found for $c_\lambda=500$ (figure \ref{fig:bifurcation}b), although in this case for oscillations from the lower static branch the time-averaged mid-point pressure is increased compared to the static solution. For some values of $c_\lambda$ we might expect the lower branch Hopf bifurcation point and the lower branch limit point to eventually merge into a co-dimension 2 (Takens-Bogdanov) bifurcation point, as previously demonstrated using lower order models of flow in collapsible tubes \citep{armitstead1996study} and channels \citep{stewart2017instabilities}. However, this possibility is not considered here.

\subsection{Homoclinic orbits} 
\label{sec:homoclinic}

\begin{figure}
\centering
\includegraphics[width=0.98\textwidth]{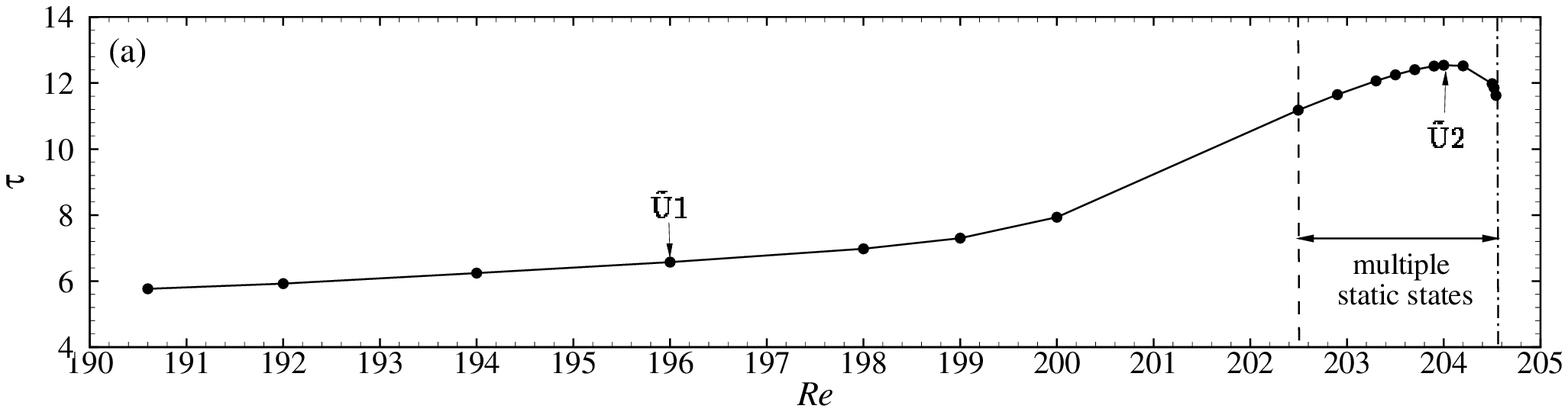}\\
\includegraphics[width=0.49\textwidth]{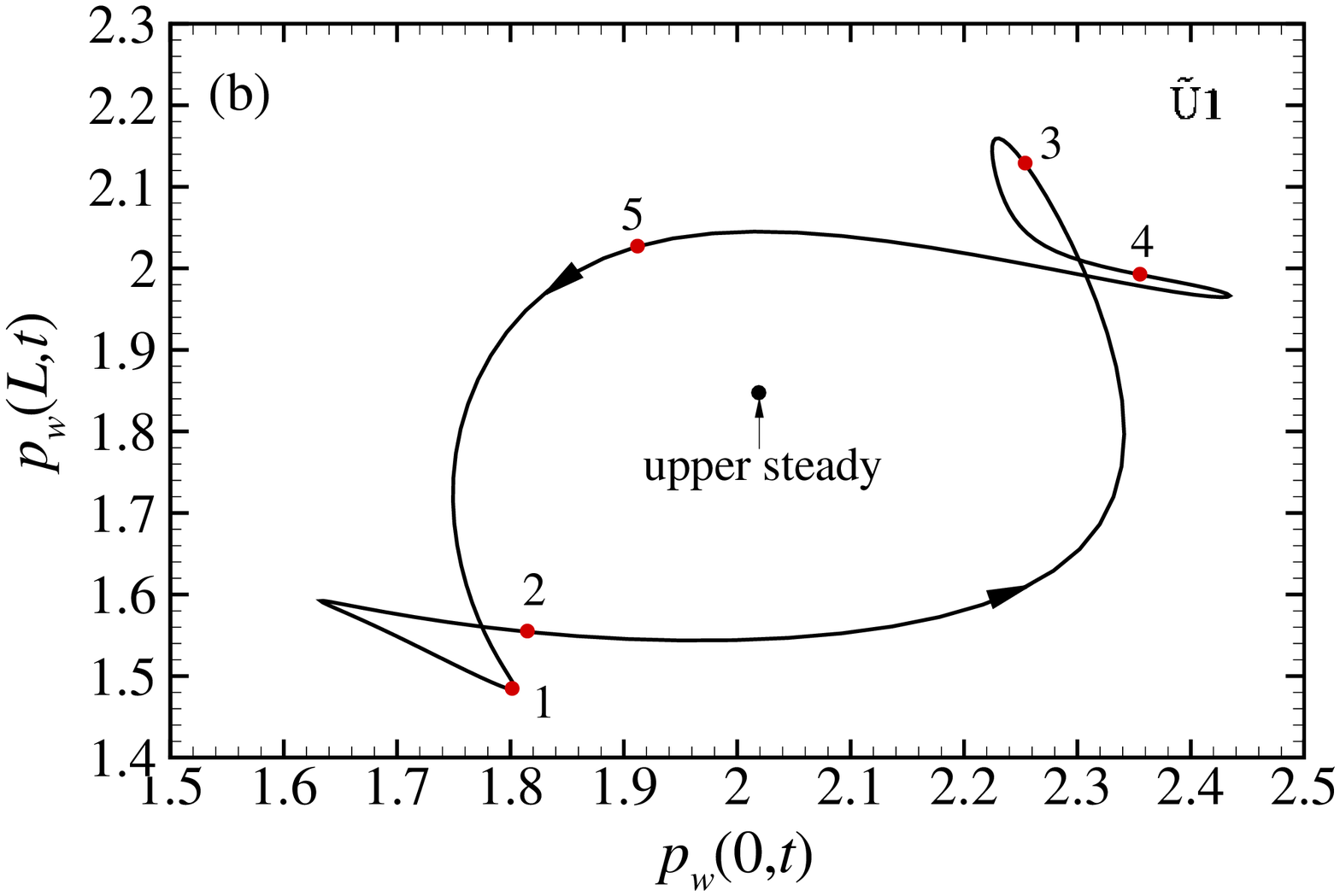}
\includegraphics[width=0.49\textwidth]{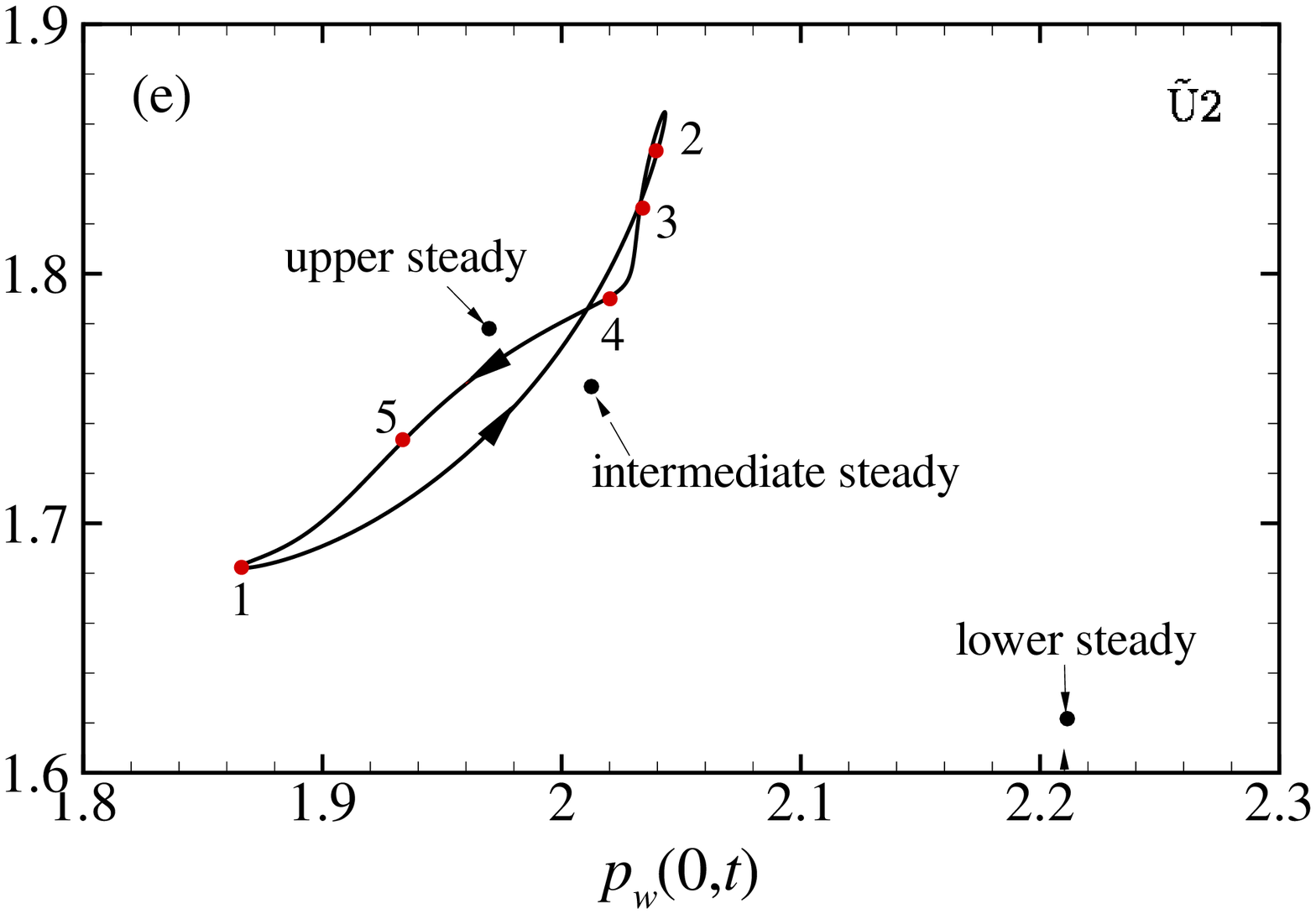}\\
\includegraphics[width=0.49\textwidth]{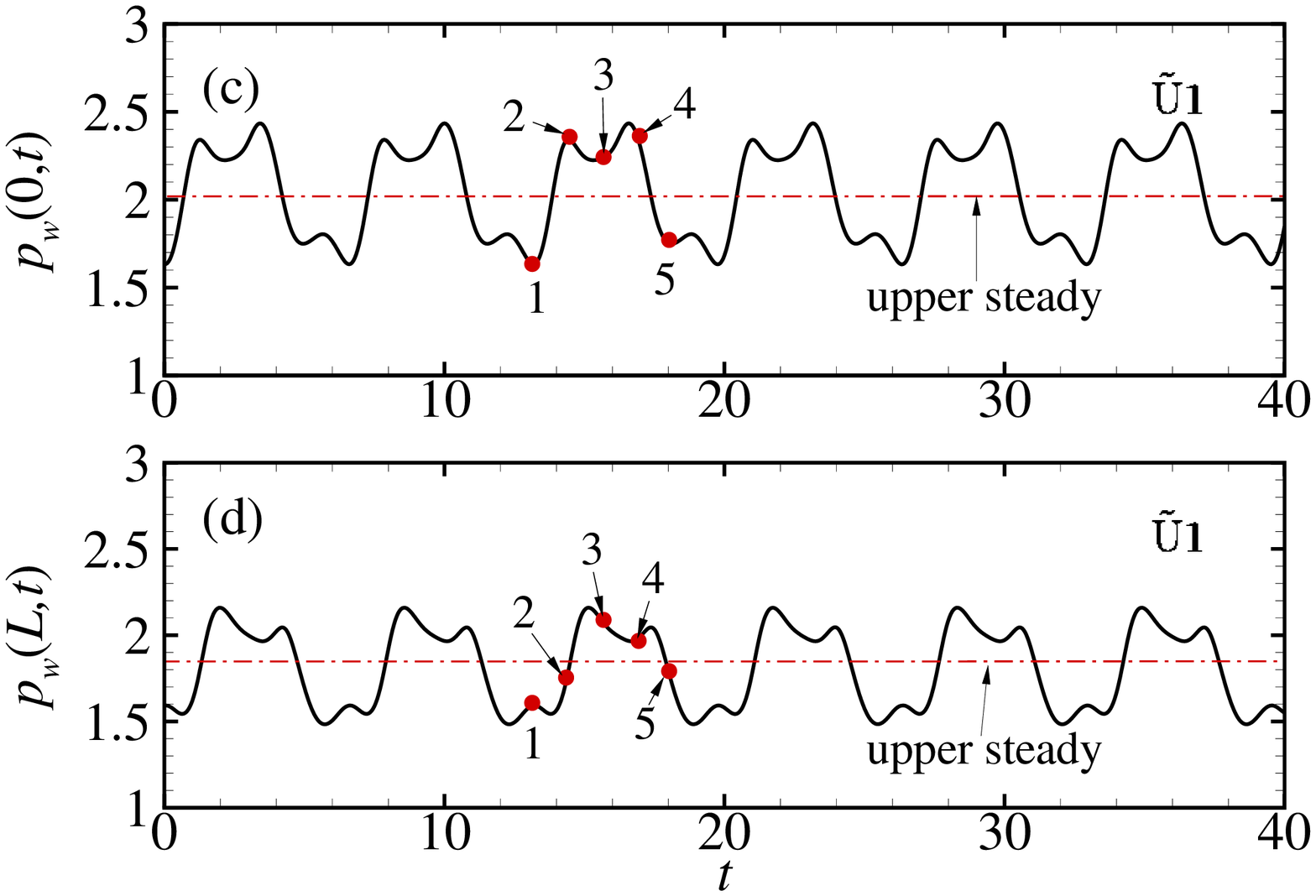}
\includegraphics[width=0.49\textwidth]{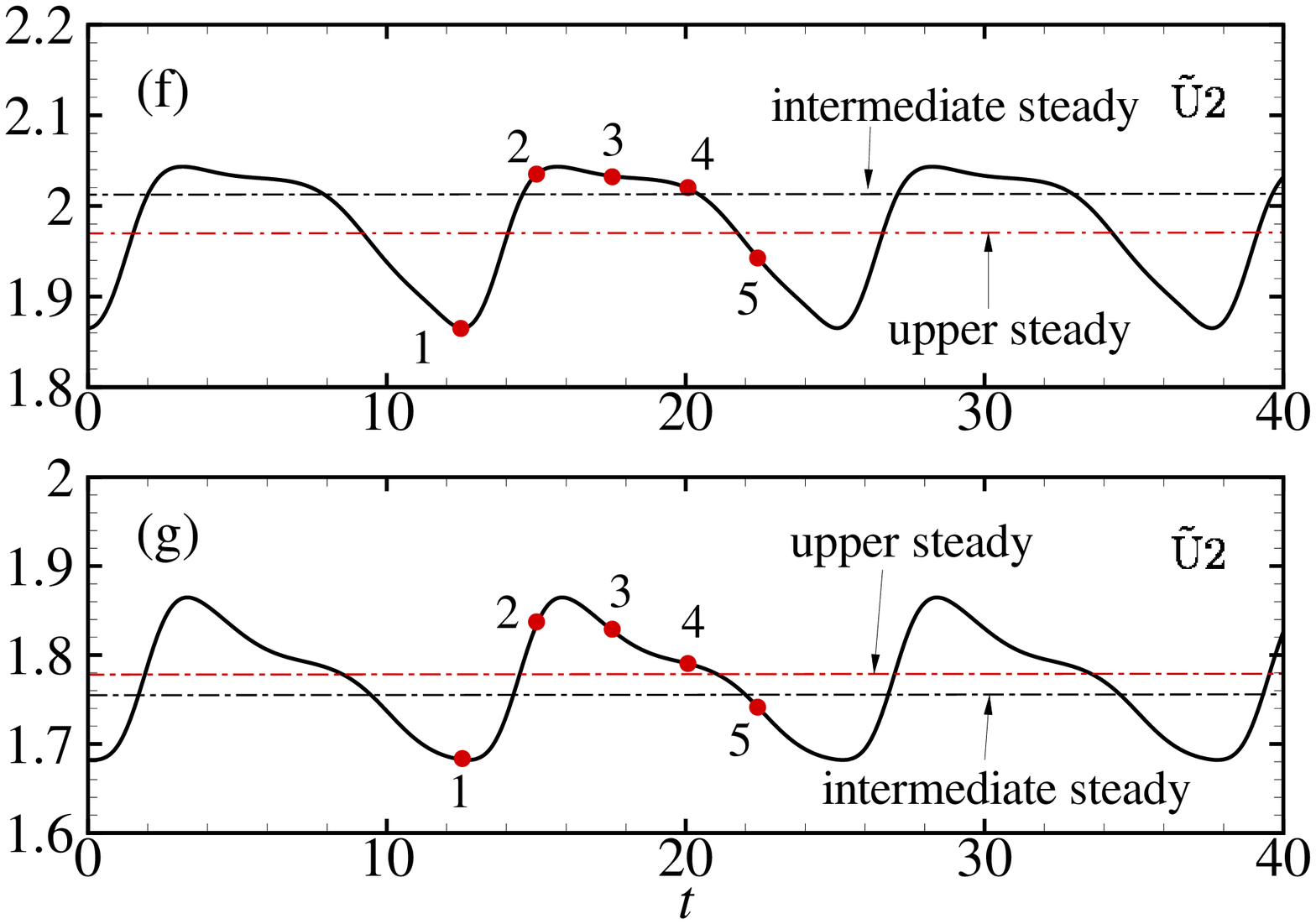}
\caption{Approach to the upper branch limit point for $c_\lambda=500$, showing: (a) the period of oscillation as a function of Reynolds number, the upper static branch enters the region with multiple static solutions at $Re\approx 202.495$ (shown with the dashed line) and terminates at the upper branch limit point $Re \approx 204.554$ (shown with the dot-dashed line); (b)  phase portrait in the space spanned by the wall pressures measured at the upstream and downstream ends of the compliant segment for operating point $\tilde{\text{U}}$1 ($Re=196$); (c) time-trace of the wall pressure at the upstream end of compliant segment for operating point $\tilde{\text{U}}$1; (d) time-trace of the wall pressure at the downstream end of compliant segment for operating point $\tilde{\text{U}}$1; (e) phase portrait in the space spanned by the wall pressures measured at the upstream and downstream ends of the compliant segment for operating point $\tilde{\text{U}}$2 ($Re=204$); (f) time-trace of the wall pressure at the upstream end of compliant segment for operating point $\tilde{\text{U}}$2; (g) time-trace of the wall pressure at the downstream end of compliant segment for operating point $\tilde{\text{U}}$2. The corresponding values of the upper, intermediate and lower static branches are marked as filled black circles in (b) and (e) and as dot-dashed lines in (c), (d), (f) and (g). Five equally-spaced time instances over a period of oscillation are marked as filled red circles in (b)-(g).}
\label{fig:wall_phase_cl500}
\end{figure}

% homoclinic orbit for clambda=500
To assess the stabilisation of the upper branch oscillations as the Reynolds number increases, in figure \ref{fig:wall_phase_cl500} we examine the interaction between the oscillatory limit cycles and the underlying static solutions in the approach to the upper branch limit point. In this case we use $c_\lambda=500$, choosing two operating points on the upper branch of static solutions denoted as $\tilde{\text{U}}$1 ($Re=196$) and $\tilde{\text{U}}$2 ($Re=204$). Note that the former is in the region where the system has a unique steady solution and the latter is in the region with multiple steady states. At onset the period of upper branch oscillation is approximately $5.76$ units; the period increases mildly as a function of Reynolds number until the amplitude reaches its maximal value (point $\tilde{\text{U}}$2). A typical phase portrait of the oscillation is illustrated in figure \ref{fig:wall_phase_cl500}(b) for operating point $\tilde{\text{U}}$1, plotted in the space spanned by the wall pressures measured at the upstream and downstream ends of the compliant segment, while the corresponding time-traces of the upstream and downstream wall pressure are shown in figures \ref{fig:wall_phase_cl500}(c) and (d), respectively. The oscillation exhibits a complicated limit cycle enclosing the corresponding upper branch static state, where the upstream and downstream pressures trace out a broadly similar sinusoidal profile, but out of phase by approximately $9\%$ of a period. Smaller amplitude pressure fluctuation are superimposed on these time-traces, so the composite profile exhibits two distinct local minima and maxima over a period; the global minimum (maximum) of the upstream pressure corresponds to a local (but not global) maximum (minimum) of the downstream pressure and vice-versa (figures \ref{fig:wall_phase_cl500}c,d), resulting in the elaborate loops in the phase portrait (figure \ref{fig:wall_phase_cl500}b). However, as the system enters the region with multiple static states, the oscillatory limit cycle shifts and no longer encloses the upper branch static state, instead becoming entrained between the upper and intermediate static states, shown for operating point $\tilde{\text{U}}$2 in figure \ref{fig:wall_phase_cl500}(e) in the space spanned by the wall pressures measured at the upstream and downstream ends of the compliant segment. In this case the time-traces of the upstream and downstream pressures have a very similar profile, and are almost perfectly in phase, although the upstream pressure is always significantly greater than the downstream (figures \ref{fig:wall_phase_cl500}f,g). Over a period the upstream wall pressure becomes very close to the intermediate static state (spending approximately two-fifths of the period in the close neighbourhood of this point, visible between points 2-4 in figure \ref{fig:wall_phase_cl500}f), indicating that the oscillatory limit cycle is interacting with the stable manifold associated with the intermediate static solution (figure \ref{fig:wall_phase_cl500}f). This interaction is accompanied by a dramatic increase in period of the oscillation (figure \ref{fig:wall_phase_cl500}a), suggestive of a nearby homoclinic orbit \citep{glendinning1994stability,strogatz2018nonlinear}.  The oscillation eventually reaches zero amplitude as the upper and intermediate static states coalesce, which results in a small decrease in the period close to the upper branch limit point (figure \ref{fig:wall_phase_cl500}a). In summary, this figure demonstrates that the upper branch oscillations can interact with the other static states of the system, resulting in a large increase in the period of oscillation and a possible homoclinic orbit.

\subsection{Energy budget for $c_{\lambda}=1600$}
\label{sec:energy}

\begin{table}
\begin{center}
\begin{tabular}{lccccccccccc}
	Case & $Re$ & ${\cal P}^{(0)}$ & ${\cal F}^{(0)}$ & ${\cal D}^{(0)}$ & ${\cal P}^{(avg)}$ & ${\cal F}^{(avg)}$ & ${\cal D}^{(avg)}$ & ${\cal P}^{(e)}$ & ${\cal F}^{(e)}$ & ${\cal D}^{(e)}$ \\[3pt]
	\hline
	U1 & $192.21$ & 2.4329 & -2.3385e-7 & 2.4329 & 2.4330 & -3.7866e-7 & 2.4330 & 1.1040e-4 & -1.4481e-7 & 9.6458e-5   \\		
	U2 & $193.0$  & 2.4240 & -2.4076e-7  & 2.4240 & 2.4269 & 6.3954e-6 & 2.4266 & 2.9312e-3 & 6.6362e-6 & 2.6277e-3 \\	
	U3 & $195.0$  & 2.4019 & -2.5682e-7  & 2.4019 & 2.4092 & 1.9983e-5 & 2.4088 & 7.3053e-3 & 2.0240e-5 & 6.8530e-3 \\	
	U4 & $200.0$  & 2.3524 & -2.6586e-7 & 2.3524 & 2.3818 & 3.6608e-5 & 2.3810 & 2.9357e-2 & 3.6874e-5 & 2.8540e-2   \\
	U5 & $201.5$  & 2.3422 & -2.2861e-7  & 2.3422 & 2.3467 & 1.4801e-5 & 2.3466 & 4.5157e-3 &  1.5029e-5 &  4.3740e-3 \\	
	U6 & $201.9$  & 2.3414 & -2.0066e-7  & 2.3414 & 2.3439 & 4.6201e-6 & 2.3439 & 2.4998e-3 & 4.8207e-6 &  2.4702e-3 \\			
	U7 & $202.1$  & 2.3422 & -1.7535e-7 & 2.3422 & 2.3439 & 2.5782e-6 & 2.3439 & 1.6730e-3 & 2.7536e-6 & 1.6424e-3 \\			
	U8 & $202.3$  & 2.3473 & -1.1288e-7  & 2.3474 & 2.3478 & -1.6674e-6 & 2.3478 & 4.3968e-4 & -1.5545e-6 & 4.3883e-4  \\
	\hline			
	I1 & $201.9$ & 2.3900 & 1.8888e-7 & 2.3900 & -& -& -& -& -&- \\	
	I2 & $202.1$ & 2.3725 & 7.4822e-8  & 2.3725 & -&- &- &- &- &- \\
	\hline			
	L1 & $201.9$  & 2.4141 & 3.6293e-7  & 2.4142 & -& -& -& -& -& -\\% 2.4141 &  3.5903e-7 & 2.4142 & 1.0807e-7 & -3.9016e-9 & -1.1522e-7 \\
	L2 & $202.1$  & 2.4224 & 4.4072e-7  & 2.4224 & -& -& -& -& -& -\\% 2.4224 & 4.4070e-7 & 2.4224 & -2.3165e-7 & -2.7203e-11 & -5.3409e-7 \\
	L3 & $202.3$  & 2.4264 & 4.9004e-7  & 2.4265 & -& -& -&- & -& -\\% 2.4264 & 4.8929e-7 & 2.4265 & -2.3702e-7 & -7.5336e-10 & 3.8717e-7  \\
	L4 & $212.0$  & 2.3902 & 1.6437e-6  & 2.3902 & -& -& -& -& -& -\\%2.3902 & 1.6502e-6 & 2.3902 & 8.1559e-8 & 6.5272e-9 &  -7.8769e-8 \\
	L5 & $212.71$  & 2.3851 & 1.7342e-6  & 2.3851 & 2.3869 & 2.2566e-5 & 2.3869 & 1.8411e-3 & 2.0832e-5 & 1.7875e-3  \\
	L6 & $213.0$  & 2.3830 & 1.7720e-6 & 2.3830 & 2.3862 & 3.1206e-5 & 2.3861 & 3.2595e-3 & 2.9434e-5 & 3.0677e-3 \\	
	L7 & $214.0$ & 2.3756 & 1.9059e-6  & 2.3756  & 2.3805 & 5.3340e-5 & 2.3804 & 4.9114e-3 & 5.1434e-5 & 4.7120e-3  \\
	L8 & $216.0$  & 2.3606 & 2.1913e-6 & 2.3607 & 2.3675 & 4.9169e-5 & 2.3669 & 6.8630e-3 & 4.6978e-5 & 6.2704e-3  \\
\end{tabular}
\caption{Computed terms in the static, time-averaged and excess energy budgets, considering for 8 operating points on the upper static branch (U1-U8), 2 points on the intermediate static branch (I1-I2) and 8 points on the lower static branch (L1-L8). Points with a hyphen do not exhibit an oscillatory limit cycle.}
\label{Tab:energy}  
\end{center}
\end{table}

% energy terms
To assess the mechanism of instability driving the self-excited oscillations, in table \ref{Tab:energy} we compute the terms in the energy budget (\ref{FulSysEnerBug:0}) at eighteen selected points with fixed $c_{\lambda}=1600$ (those labelled U1-U8, I1-I2, L1-L8 in figure \ref{Fig:st-ymin-max}). In particular, we compute the energy budget for the different branches of self-excited oscillation by post-processing the fully developed limit cycles and taking the time-average over a period (according to Eq.~(\ref{eq:timeaverage})); these terms are denoted with the superscript $^{(avg)}$. We further compute the terms in the energy budget for the corresponding static states (Eq.~(\ref{B_FullEner:st})), denoted with the superscript $^{(0)}$, and the excess between the time-averaged and the static (Eq.~(\ref{B_FullEner:Osci})), denoted with the superscript $^{(e)}$. 

% static energies
The contributions to the static energy budget can be computed at all the operating points, where the work done by upstream pressure (${\cal P}^{(0)}$) almost exactly balances the work done by dissipation in the fluid (${\cal D}^{(0)}$) in every case, while the kinetic energy flux (${\cal F}^{(0)}$) is negligible in comparison. Such a result is to be expected because in a steady configuration the outlet flux must exactly balance the prescribed inlet flux. We find that the work done by upstream pressure generally decreases with increasing Reynolds number for all three branches. However, for fixed Reynolds number (in the region with multiple steady states), the lower static branch requires more work to be done by the driving pressure than the intermediate static branch, which in turn requires more work to be done by the driving pressure than the upper static branch.

% oscillatory energies
For fully developed limit cycles we observe that the dominant energy balance is always between the rate of working of upstream pressure and the rate of working of viscous dissipation, similar to previous studies of oscillations driven by fixed upstream flux \citep{stewart2017instabilities}. Again, we might expect the net kinetic energy flux extracted from the mean flow $({\cal F}^{(avg)})$ to be very small since the time-averaged flux at the outlet boundary must be unity to conserve mass over a period.

% excess energies
For those operating points which exhibit a fully developed limit cycle we can further compute the excess energy compared to the static, where we find that the excess work done by upstream pressure is positive (${\cal P}^{(e)}>0$) and so the upstream pressure must work harder to sustain the oscillation; this increase is balanced by an increase in the work done by dissipation from the more complicated flow-field (${\cal D}^{(e)}$). This increase in working of upstream pressure is presumably achieved by the action nonlinear Reynolds stresses. Hence, this fluid-beam system shows a very different mechanism of energy transfer to that of the fluid-membrane system described by \citet{stewart2017instabilities}, who found that onset of oscillation reduces the overall dissipation energy by increasing the time-averaged minimal width of the channel. Thus, in that case the upstream pressure works less hard to drive the instability compared to the steady configuration. However, oscillations arising due to an increase in the rate of working of upstream pressure have been discussed elsewhere in the literature \citep{jensen2003high,stewart2009local,stewart2010sloshing}.

\section{Comparison to the previous fluid-beam model}
\label{sec:compare}

\begin{figure}
\centering
\includegraphics[width=0.98\textwidth]{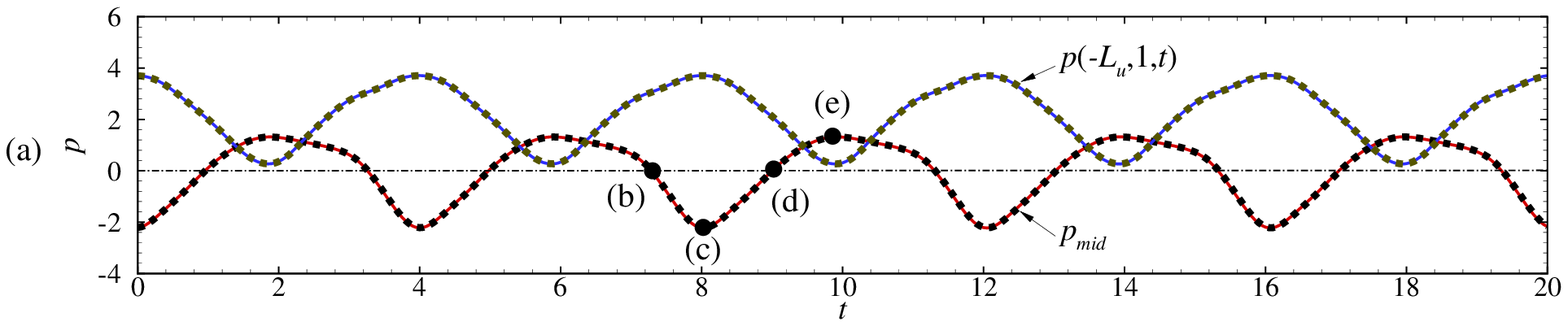}\\
\includegraphics[width=0.98\textwidth]{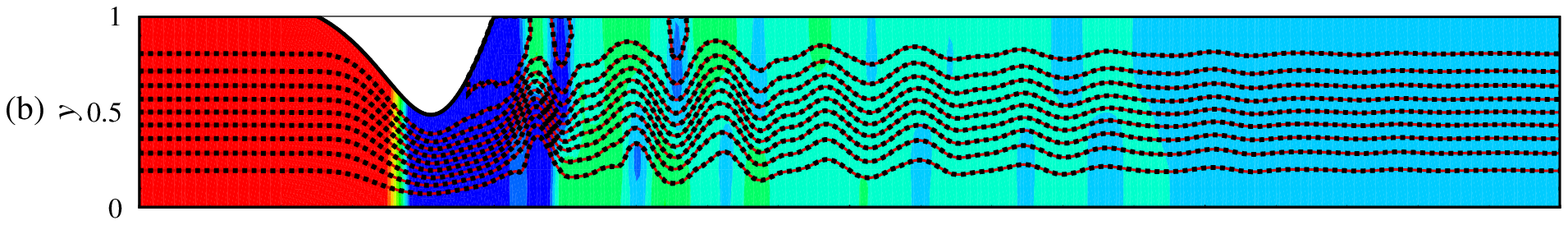}\\
\includegraphics[width=0.98\textwidth]{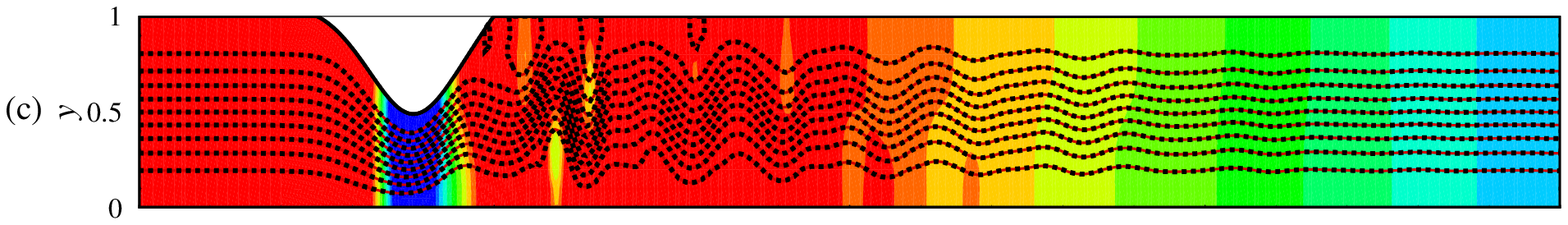}\\
\includegraphics[width=0.98\textwidth]{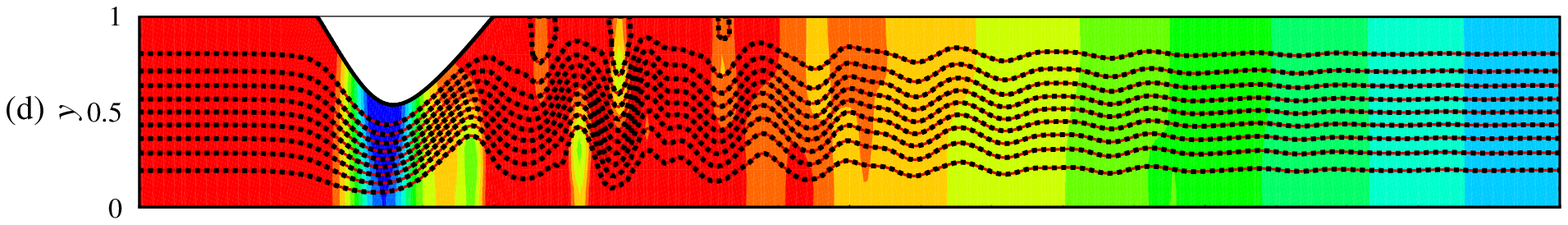}\\
\includegraphics[width=0.98\textwidth]{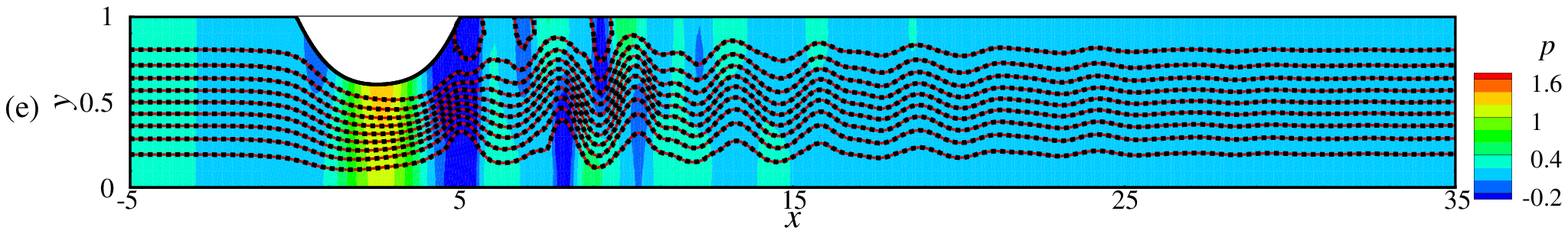}
\caption{Comparison of the fluid-beam model using the `old' and `new' Kirchhoff laws (Eq.~(\ref{KirLaw-old}) and Eq.~(\ref{KirLaw}), respectively) at operating point $\tilde{\text{L}}$1 ($c_\lambda=500$, $Re=400$) showing: (a) time-traces of the upstream driving pressure and the wall midpoint pressure using the new (solid lines) and old (dashed lines) Kirchhoff laws, respectively, and (b)-(e): four snapshots of the fluid streamlines for the new (red solid lines) and old (black dashed lines) Kirchhoff laws, superimposed on pressure contours within the collapsible channel for the new Kirchhoff law illustrated with a colour contour map. The dot-dashed line in panel (a) is the steady wall mid-point pressure using the new Kirchhoff law.}
\label{Fig:U10-nkl-okl}
\end{figure}

In order to assess how our modifications to the Kirchhoff law influence the predictions of the fluid-beam model, in figure \ref{Fig:U10-nkl-okl} we consider operating point $\tilde{\text{L}}$1 ($c_\lambda=500$, $Re=400$ along the lower static branch) and compare fully developed limit cycles found using our constitutive law (Eq.~(\ref{KirLaw}), which we term the `new' model) to predictions using the constitutive law of \citet{luo2008cascade} (Eq.~(\ref{KirLaw-old}), which we term the `old' model). We choose this operating point at the largest value of Reynolds number considered to maximise the deformation of the beam (see figure \ref{fig:bifurcation}). In particular, we illustrate time-traces of the upstream driving pressure measure $p(-L_u,1,t)$ and the wall midpoint pressure $p_{mid}$ computed using these two models (figure \ref{Fig:U10-nkl-okl}a) and compare the resulting pressure contours at four time instances (when $p_{mid}$ reaches zero and its extrema, respectively) over a period (figures \ref{Fig:U10-nkl-okl}b-e). In all cases we find that the two flows are almost indistinguishable, so the modification made to the wall model to make it energetically conservative makes negligible difference to the predictions. For this oscillation arising from the lower static branch (see figure \ref{fig:bifurcation}) the channel wall is highly collapsed over the entire period of oscillation, shedding large amplitude vorticity waves into the downstream rigid segment \cite[][]{stephanoff1983fluid}. Over a period, the oscillation proceeds by altering the wall shape to constrict the channel (relative to the static) and develop a wide region of low pressure (less than the channel outlet pressure) at the downstream end of the flexible segment (figure \ref{Fig:U10-nkl-okl}b). This pressure difference locally reverses the flow, which gradually propagates the channel constriction upstream (figure \ref{Fig:U10-nkl-okl}c). The upstream driving pressure must then work harder to sustain the prescribed flow (figure \ref{Fig:U10-nkl-okl}a), while the region of low pressure becomes narrower (figure \ref{Fig:U10-nkl-okl}c). Due to local flow reversal close to the constriction, conservation of mass dictates that the wall must expand upstream to accommodate the incoming fluid. However, resistance of the wall to bending and stretching prevents its continued expansion and so drives the constriction downstream again (\ref{Fig:U10-nkl-okl}a,d), lowering the driving pressure and moving the region of low pressure toward the downstream end of the flexible segment (figure \ref{Fig:U10-nkl-okl}e), and the process repeats. In this case the $x$-position of the channel constriction changes smoothly as there is one local minimum width for all time. In summary, this figure demonstrates that our modification to the Kirchhoff law makes negligible difference to the fully developed oscillations along the lower branch. It also provides insight into the mechanism of instability driving the lower branch oscillation, where the channel constriction smoothly propagates backwards and forwards along the compliant segment.

\section{Discussion}

% overview of the model, assumptions, limitations
This study considered a model for flow through a finite length collapsible channel driven by a fixed upstream flux, where the flexible wall takes the form of a long thin elastic beam with a modified Kirchhoff law, slightly adjusting the formulation of \citet{luo2008cascade} to ensure that the wall model is energetically conservative. However, this modification makes almost no difference to the final predictions (figure \ref{Fig:U10-nkl-okl}). The full unsteady model was solved using an Arbitrary-Lagrangian-Eulerian framework based on the finite element method.

% static
The model predicts that there is always at least one static solution for all parameters tested and across most of the parameter space this static profile is unique. However, for sufficiently large Reynolds numbers there exists regions of parameter space with three co-existing static states (figure \ref{Fig:st-ymin-max}): an upper branch (where the channel is almost entirely inflated), a lower branch (where the wall is mostly collapsed) with an intermediate branch connected by a pair of limit point bifurcations. Such three branch behaviour has been previously reported in collapsible tube experiments \cite[][]{bertram1991application} and computations \cite[][]{heil2010self}, and has also been found in flow through collapsible channels where the elastic wall has large pre-stress (e.g. as a thin membrane \citep{luo2000multiple,stewart2017instabilities}, using nonlinear shell theory \citep{heil2004efficient} or as a hyperelastic material of finite thickness \citep{herrada2021global}), but the present work is the first time it has been demonstrated in a fluid-beam model with no pre-stress but with resistance to both bending and stretching.

% unsteady predictions, lower branch
In line with previous studies \citep{luo2008cascade,hao2016arnoldi}, the lower branch of static solutions becomes unstable to self-excited oscillations via a supercritical Hopf bifurcation when the Reynolds number exceeds a critical threshold (figure \ref{fig:bifurcation}, where typical flow profiles over a period of oscillation are shown in figure \ref{Fig:U10-nkl-okl});  analogous oscillations growing from the lower static branch have already been reported elsewhere \citep{heil2004efficient,stewart2017instabilities,herrada2021global}. Over a period of these oscillations the profile of the flexible wall exhibits a single minimum which smoothly propagates up and downstream. However, it emerges that the cascade structure described by \citet{luo2008cascade} is by no means exhaustive, showing that there are other neutral stability curves forming unstable islands in zones which were previously deemed stable (figure \ref{Fig:st-saddle-neutral}). 

% unsteady predictions, upper branch
In particular, our simulations demonstrate that the upper branch of static solutions can also become unstable to oscillations, and this transition typically occurs for lower Reynolds numbers that those considered by \citet{luo2008cascade}. However, it should be noted that this upper branch instability is not a consequence of our modification to the wall model - subsequent analysis using the original model of \citet{luo2008cascade} has confirmed the existence of this upper branch instability. A similar instability of the upper static branch has very recently been reported by \citet{herrada2021global}, using a global linear stability eigensolver. However, our method provides access to the oscillatory limit cycle, where we showed that the upper branch instability develops into a oscillation where the wall is almost entirely inflated over a period, growing an outward bulge at the downstream end of the domain which propagates upstream and is eventually suppressed by interaction with flow in the upstream rigid segment. The saturated amplitude of these oscillations initially grows with Reynolds number but eventually reaches a maximum and then decreases, terminating at the end of the upper branch of static solutions (the upper branch limit point). This demise is accompanied by a significant increase in the period of oscillation as the nonlinear limit cycles exhibit strong interaction with the intermediate static state, reminiscent of a homoclinic orbit (figure \ref{fig:wall_phase_cl500}). Stabilisation of the upper branch results in a range of Reynolds numbers beyond this limit point where the system is entirely stable, before eventually becoming unstable to oscillations growing from the lower static branch. Hence, the cascade structure observed in this system is at least partially due to the complexity of the underlying static state.

% energy
Our modified fluid-beam formulation means that the elastic wall is perfectly energetically conservative, and so over a period of oscillation the only non-trivial contributions to the energy budget are the work done by upstream pressure, the work done by viscous dissipation and the net kinetic energy flux extracted from the mean flow. Of these, the latter is negligible as the upstream flux is prescribed and the oscillation must conserve mass over a period (see table \ref{Tab:energy}), similar to observations made using lower order models \citep{xu2015low,stewart2017instabilities}. For oscillations arising from both the upper and lower static branches, the work done by upstream pressure is increased by the oscillation compared to its corresponding static value. Hence, the system must work harder to oscillate, with the extra energy supplied by the action of nonlinear Reynolds stresses \citep[a similar mechanism for pressure-driven oscillations from a non-uniform basic state was previously reported by][]{stewart2009local}. This mechanism is different to that reported for lower branch oscillations when the external pressure is very large \citep{stewart2017instabilities}, where there the oscillation caused the collapsed channel to expand slightly, reducing the overall viscous dissipation. The energy analysis presented in this paper provides insight into the fundamental mechanism of instability in this system. Although simplified, this model can also serve as a mathematical prototype for a general fluid-structure interaction systems commonly found in clinical applications with periodic motion, such as blood flows in the aorta and large coronary arteries, and oscillatory airflows in lung airways.  
 
\section*{Acknowledgements}

We gratefully acknowledge funding from the Chinese Scholarship Council (DYW), UK Engineering and Physical Sciences Research Council grants EP/S020950,   EP/S030875, and EP/N014642 (XYL and PSS).  Special thanks to Prof.~Zongxi Cai (Tianjin University, China) for helpful discussions. 

\section*{Declaration of interests}

The authors report no conflict of interest.

\section*{Data Accessibility}

The data presented in this paper is accessible at \\ {\tt http://dx.doi.org/10.5525/gla.researchdata.1112}.

\appendix

\section{Rate of working of external pressure}
\label{appA}

In this appendix we manipulate the rate of working of external pressure (Eq.~(\ref{eq:extpress})) to be expressed as a complete time derivative.

The relationships between the Eulerian and Lagrangian coordinate systems take the form \citep{cai2003fluid},
\begin{align}
\mathbf e_1=\frac{1}{\lambda}\left(\frac{\partial x_b}{\partial l}{\mathbf g_1}+\frac{\partial y_b}{\partial l}{\mathbf g_2}\right),\quad \mathbf e_2=\frac{1}{\lambda}\left(-\frac{\partial y_b}{\partial l}{\mathbf g_1}+\frac{\partial x_b}{\partial l}{\mathbf g_2}\right), \quad \mathbf e_3=\mathbf g_3.\label{e1e2}
\end{align}
Hence, substituting the velocity of the beam
\begin{equation}
{\mathbf u}_b=\frac{\partial{\mathbf x}_b}{\partial t} = \frac{\partial x_b}{\partial t}\mathbf g_1 + \frac{\partial y_b}{\partial t} \mathbf g_2,
\end{equation}
and the unit vector $\mathbf e_2$ (\ref{e1e2}) into the rate of working of external pressure ${\cal P}_e$ (\ref{eq:extpress}) we obtain
\begin{align}
{\cal P}_e&=\int_0^L -p_e\left(\frac{\partial x_b}{\partial l}\frac{\partial y_b}{\partial t}-\frac{\partial x_b}{\partial t}\frac{\partial y_b}{\partial l}\right)\,{\rm d}l=\frac{\partial}{\partial t}\int_0^L-p_e \frac{\partial x_b}{\partial l}y_b\,{\rm d}l+\left[p_e\frac{\partial x_b}{\partial t}y_b\right]_0^L.\label{d:pe}
\end{align}
The second term vanishes after applying the boundary condition (\ref{BBodCon:1}). Hence, the rate of working of external pressure ${\cal P}_e$ can be written as a complete time derivative in the form presented in Eq.~(\ref{eq:extpress}).

\section{Convergence of the numerical method}
\label{sec:appB}

\begin{figure}
\centering
\includegraphics[width=0.98\textwidth]{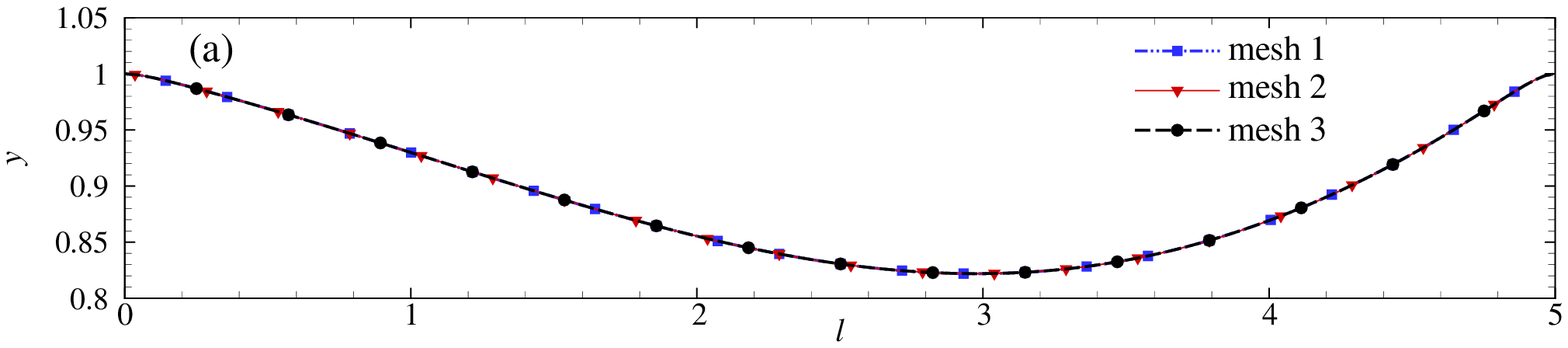}
\includegraphics[width=0.98\textwidth]{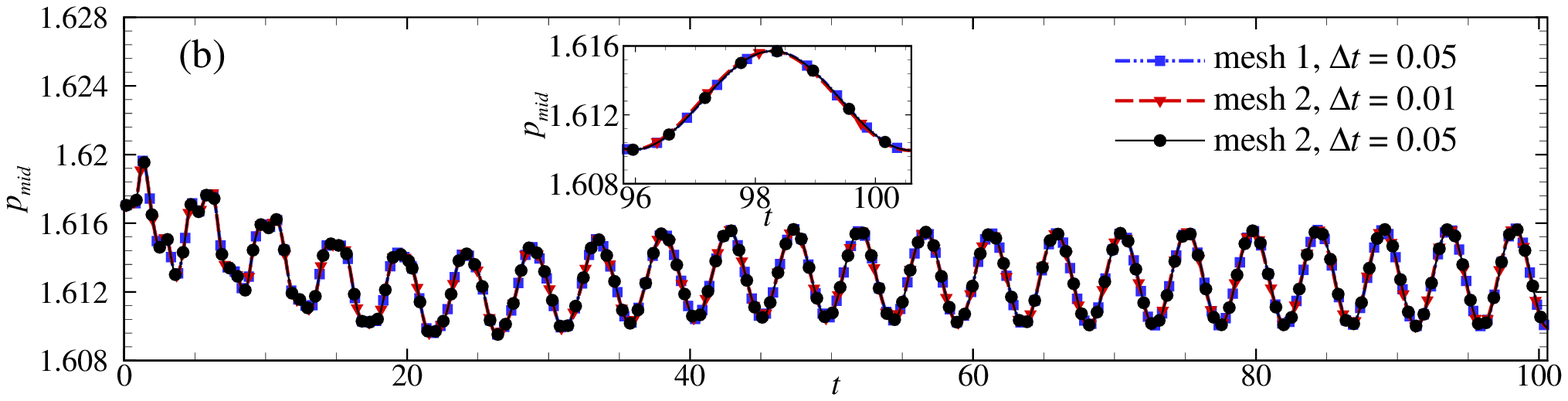}
\caption{Convergence of the numerical method illustrated at operating point L8 ($Re=216$, $c_{\lambda}=1600$): (a) the static beam profile computed using mesh 1 (filled squares), mesh 2 (inverted triangles) and mesh 3 (filled circles); (b) time-trace of the fluid pressure at the beam mid-point $p_{mid}(t)$ ($x=L/2$) for three different combinations of mesh and timestep, including mesh 1 with $\Delta t=0.05$ (filled squares), mesh 2 with $\Delta t=0.01$ (inverted triangles) and mesh 2 with $\Delta t=0.05$ (filled circles).}
\label{Fig:mesh-validation}
\end{figure}

The convergence of the numerical method is illustrated in figure \ref{Fig:mesh-validation} for operating point L8 ($Re=216$, $c_{\lambda}=1600$). In particular, we consider simulations with three finite element meshes composed of either 29697 elements in the bulk and 120 elements along the beam (which we term mesh 1), 36657 elements in the bulk and 140 elements along the beam (which we term mesh 2) or 41877 elements in the bulk and 140 elements along the beam (which we term mesh 3). The static beam shape computed using these three meshes is shown in figure \ref{Fig:mesh-validation}(a), where we find that the profile is indistinguishable across the three cases. The static simulations in the main text were all computed using mesh 2. Furthermore, we consider unsteady simulations using meshes 1 and 2 for a variety of choices of timestep (figure \ref{Fig:mesh-validation}b) for identical initial conditions. Note that the mesh is regenerated after each timestep (according to the procedure discussed in the main text), but the number of elements stays fixed. We find that these time-traces are indistinguishable across the three different choices, indicating that our simulations are well converged. The unsteady simulations in the main text were all computed using mesh 2 with a timestep $\Delta t=0.01$. This ensures at least 500 timesteps per period of oscillation for the examples considered in this paper.

\bibliographystyle{jfm}
\bibliography{ref}
%Use of the above commands will create a bibliography using the .bib file. Shown below is a bibliography built from individual items.

\end{document}